%% file: mindmansion.tex
\documentclass[sigconf, authorversion]{acmart}

\AtBeginDocument{%
  \providecommand\BibTeX{{%
    \normalfont B\kern-0.5em{\scshape i\kern-0.25em b}\kern-0.8em\TeX}}}

\input{meta/packages}

\input{meta/acronyms}
\input{meta/definitions}

\input{meta/stats}
\input{meta/quotes}

\copyrightyear{2024}
\acmYear{2024}
\setcopyright{acmlicensed}\acmConference[DIS '24]{Designing Interactive Systems Conference}{July 1--5, 2024}{IT University of Copenhagen, Denmark}
\acmBooktitle{Designing Interactive Systems Conference (DIS '24), July 1--5, 2024, IT University of Copenhagen, Denmark}
\acmDOI{10.1145/3643834.3661557}
\acmISBN{979-8-4007-0583-0/24/07}

\begin{document}

\title[\systemname{}]{\systemname{}: \longtitle{}}

\input{meta/authors}

\input{content/00_abstract}

\input{meta/meta}

\input{img/teaser}

\maketitle

\input{content/01_introduction}
\input{content/02_relatedwork}

\input{content/03_concept}

\input{content/04_methodology}

\input{content/05_results}

\input{content/06_discussion}
\input{content/07_limitations}
\input{content/08_conclusion}

\bibliographystyle{ACM-Reference-Format}
\bibliography{references}

\end{document}

%% file: meta/packages.tex
\usepackage{todonotes}
\usepackage[nolist,nohyperlinks ]{acronym}
\usepackage{siunitx}
\usepackage{calculator}
\usepackage{csquotes}
\usepackage{subcaption} 
\usepackage{fmtcount}
\usepackage{enumitem}
\setlist[description]{%
labelindent=0.5\parindent,%
itemindent=-.6em,%
leftmargin=*,%
}

%% file: meta/acronyms.tex
\begin{acronym}[UMLX]
	\acro{HMD}{head-mounted display}
	\acro{GUI}{graphical user interface}
 	\acro{UI}{user interface}
	\acro{HUD}{head-up display}
	\acro{TCT}{task-completion time}
	\acro{SVM}{support vector machine}
	\acro{EMM}{estimated marginal mean}
	\acro{AR}{Augmented Reality}
 	\acro{MR}{Mixed Reality}
	\acro{VR}{Virtual Reality}
    \acro{VE}{Virtual Environment}
	\acro{CSCW}{Computer-Supported Cooperative Work}
	\acro{HCI}{Human-Computer Interaction}
	\acro{ABI}{Around-Body Interaction}
	\acro{SLAM}{Simultaneous Location and Mapping}
	\acro{ROM}{range of motion}
	\acro{EMG}{electromyography}
	\acro{STS}{System Trust Scale}
	\acro{IPQ}{Igroup Presence Questionnaire}
	\acro{TLX}{NASA Task Load Index}
 	\acro{RTLX}{Raw Nasa-TLX}
 	\acro{CAD}{Computer Aided Design}
    \acro{ART}{Aligned Rank Transform}
	\acro{IOS}{Inclusion of Other in the Self}
	\acro{GEQ}{Game Experience Questionnaire}
    \acro{PANAS}{Positive and Negative Affect Schedule}
    \acro{CBT}{Cognitive Behavioral Therapy}
    \acro{ACT}{Acceptance and Commitment Therapy}
    \acro{ATQ}{Automatic Thoughts Questionnaire}
\end{acronym}

%% file: meta/definitions.tex
\newcommand{\systemname}{Mind Mansion}
\newcommand{\longtitle}{Exploring Metaphorical Interactions to Engage with Negative Thoughts in Virtual Reality}

\newcommand{\ivMetaphoricalInteraction}{\textsc{Metaphorical Interaction}}

\newcommand{\ivBottleRefilling}{\textsc{BottleRefilling}}
\newcommand{\ivWateringPlants}{\textsc{WateringPlants}}
\newcommand{\ivBurstingBalloons}{\textsc{BurstingBalloons}}
\newcommand{\ivRisingBalloons}{\textsc{RisingBalloons}}
\newcommand{\ivCarpetBeating}{\textsc{CarpetBeating}}
\newcommand{\ivPillowPunching}{\textsc{PillowPunching}}
\newcommand{\ivPuddleSweeping}{\textsc{PuddleSweeping}}
\newcommand{\ivWindowWiping}{\textsc{WindowWiping}}
\newcommand{\ivTrashbagThrowing}{\textsc{TrashbagThrowing}}
\newcommand{\ivBookshelfSorting}{\textsc{BookshelfSorting}}

\newcommand{\dvCoping}{\textsc{\textsc{Coping}}}
\newcommand{\dvEngagement}{\textsc{\textsc{Engagement}}}

%% file: meta/stats.tex
\newcommand{\ano}[4]{$F_{#1, #2}=#3$, $p#4$}

\newcommand{\subEtaG}[2]{%
	\ifthenelse{\equal{#1}{\string >.05}}
	{}
	{, $\eta_{G}^{2}=#2$}%
}

\newcommand{\subEta}[2]{%
	\ifthenelse{\equal{#1}{\string >.05}}
	{}
	{, $\eta^{2}=#2$}%
}

\def\ges{$\eta_{G}^{2}$}

\newcommand{\efETAsquared}[1]{%
	\ifdim#1pt>0.139pt 
	large (\ges{} = #1)
	\else 
	\ifdim#1pt>0.059pt 
	medium (\ges{} = #1)
	\else 
	small (\ges{} = #1)
	\fi
	\fi
}

%% file: meta/quotes.tex
\usepackage{etoolbox}

%% file: meta/authors.tex

\author{Julian Rasch}
\orcid{0000-0002-9981-6952}
\affiliation{
  \institution{LMU Munich}
  \streetaddress{Frauenlobstr. 7A}
  \city{Munich}
  \country{Germany}
  \postcode{80337}
}
\email{julian.rasch@ifi.lmu.de}

\author{Michelle Johanna Zender}
\orcid{0009-0002-2379-0394}
\affiliation{
  \institution{LMU Munich}
  \streetaddress{Frauenlobstr. 7A}
  \city{Munich}
  \country{Germany}
  \postcode{80337}
}
\email{michelle.zender@campus.lmu.de}

\author{Sophia Sakel}
\orcid{0000-0002-6326-8018}
\affiliation{
 \institution{LMU Munich}
  \streetaddress{Frauenlobstr. 7A}
  \city{Munich}
  \country{Germany}
  \postcode{80337}
}
\email{sophia.sakel@ifi.lmu.de}

\author{Nadine Wagener}
\orcid{0000-0003-4572-4646}
\affiliation{
  \institution{University of Bremen}
  \streetaddress{}
  \city{Bremen}
  \country{Germany}
  \postcode{}
}
\email{nwagener@uni-bremen.de}

\renewcommand{\shortauthors}{Rasch et al.}

%% file: content/00_abstract.tex

\begin{abstract}

Recurrent negative thoughts can significantly disrupt daily life and contribute to negative emotional states. Facing, confronting, and noticing such thoughts without support can be challenging. To provide a playful setting and leverage the technical maturation of \ac{VR}, our \ac{VR} experience, \textsc{Mind~Mansion}, places the user in an initially cluttered virtual apartment. Here we utilize established concepts from traditional therapy and metaphors identified in prior works to let users engage metaphorically with representations of thoughts, gradually sorting the space, fostering awareness of thoughts, and supporting mental self-care.
The results of our user study (n = 30) reveal that \textsc{Mind~Mansion} encourages the exploration of alternative perspectives, fosters acceptance, and potentially offers new coping mechanisms. Our findings suggest that this \ac{VR} intervention can reduce negative affect and improve overall emotional awareness.

\end{abstract}


%% file: meta/meta.tex
\ccsdesc[300]{Human-centered computing~Collaborative and social computing}
\ccsdesc[500]{Human-centered computing~Virtual reality}
\ccsdesc[300]{Human-centered computing~HCI theory, concepts and models}
\ccsdesc[300]{Human-centered computing~Interaction techniques}
\keywords{Interactions, Therapeutic Applications, Mental Health, Virtual Reality}

%% file: img/teaser.tex
\begin{teaserfigure}
	\includegraphics[width=\textwidth]{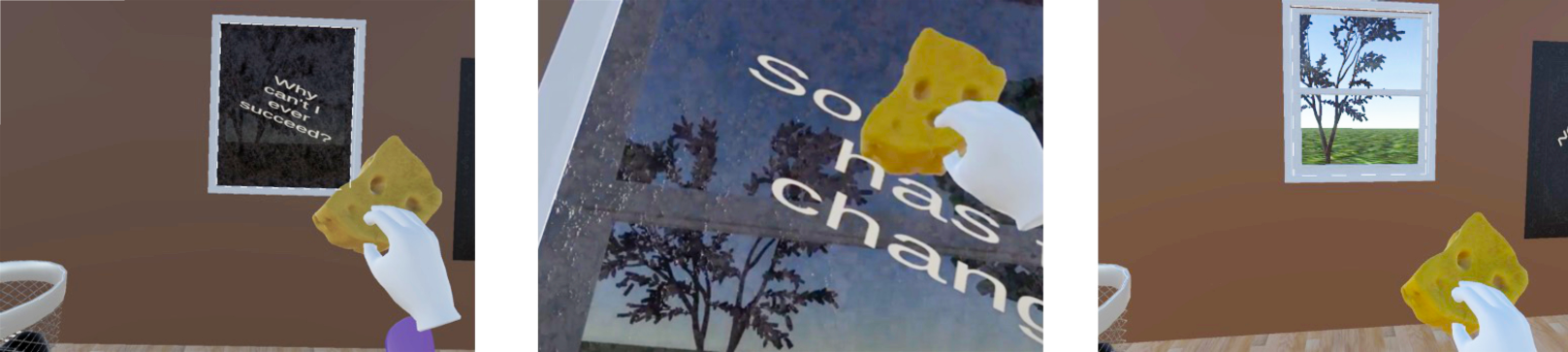}
	\caption{In this paper, we explore the potential of metaphorical interactions with negative thoughts in VR in the context of everyday interactions. In \textsc{Mind~Mansion} users gradually sort and clean their virtual apartment, leading to a more friendly environment through engagement with thoughts.}
	\Description[Teaser Figure for Mind Mansion]{This teaser figure of the paper consists of three parts. The first part shows a dark wall with a dirty window with some text on it. A hand holds a sponge. The second image shows the hand with the sponge cleaning the window. The third image shows the cleaned window without text on it. The wall and room now appear brighter and more friendly.}
	\label{fig:teaser}
\end{teaserfigure}

%% file: content/01_introduction.tex
\section{Introduction}
\label{sec:introduction}






Recurring negative thoughts trouble many people \cite{belloch_intrusive_2004, larsson_using_2016}, and can be challenging to regulate consciously, consequently hindering daily functioning and contributing to negative emotional states like increased levels of stress and anxiety~\cite{clark_unwanted_2005, eikey_beyond_2021}. Actively perceiving and facing these challenging emotions and thoughts can facilitate long-term mental well-being~\cite{gross_emotion_2015, webb_dealing_2012}.
Individuals often require external support to reflect on and question existing negative thoughts and emotions. This can encourage alternative perspectives and offer impulses for effective coping strategies. While this external support can come from literature or a social circle, traditional therapeutic interventions such as Cognitive Behavior Therapy (CBT) and Acceptance and Commitment Therapy (ACT) are common approaches in treating negative emotions~\cite{butler_empirical_2006, larsson_using_2016, garnefski_relationship_2002, hofmann_acceptance_2008}, building upon techniques like cognitive restructuring, cognitive defusion~\cite{larsson_using_2016} and emotional regulation~\cite{gross_emotion_2015}, guiding patients to critically evaluate the validity of their negative thoughts and encourage the development of alternative, more realistic thoughts that better align with their actual circumstances~\cite{larsson_using_2016}. 
Despite its effectiveness, access to these interventions or suitable conversational partners is often limited, leaving many individuals experiencing negative emotions with no access to these treatments, especially on an everyday basis~\cite {lovell_multiple_2000}.

Here, technology can offer new opportunities to support and provide coping strategies for mental health challenges, e.g., by offering individuals autonomous mental health management in their everyday routine~\cite{smith_digital_2022, wadley_digital_2020}. 
With the technical maturing and rising availability of affordable \acp{HMD}, also \ac{VR} is becoming a promising and innovative tool for frequent use, making it a powerful tool for advancing the understanding, assessment, and treatment of mental health issues. \ac{VR} allows for controlled yet immersive environments~\cite{loomis_immersive_1999}, blocking out distractions from the real world and allowing metaphoric actions that are not possible in reality (e.g.~\cite{brinol_objectification_2017, grieger_trash_2021}).
In that regard, recent research has explored VR applications and approaches to practice and explore coping strategies for negative emotions and techniques to support common therapy approaches~\cite{baghaei_time_2020}. For instance, \ac{VR} applications can help visualize emotions for emotional engagement~\cite{wagener_mood_2022} and facilitate reflecting on past challenges~\cite{wagener_selvreflect_2023}. Further, \citet{grieger_trash_2021} explored two approaches to manipulate and destroy negative text messages, while \citet{wang_break_2018} researched the potential of embodying metaphors. 
Additionally, \citet{wagener_letting_2023} identified design approaches for future VR applications addressing negative emotion states.
Despite these efforts, little research addresses the direct confrontation of negative thoughts in a \ac{VR} context.


To address this gap, we developed \textsc{Mind~Mansion}, a \ac{VR} application exploring how to directly interact and confront negative thoughts in \ac{VR}. In particular, we investigate how metaphorical interactions can be implemented and how they can assist in effectively coping with negative thoughts. Our approach aims to provide users with a way to engage with their thoughts by cleaning a virtual apartment, which serves as a visual representation of the cognitive process of sorting one’s thoughts. This approach finds support in established research on repetitive movements as a coping strategy~\cite{lang_effects_2015}, highlighting the positive effects of decluttering~\cite{saxbe_no_2010} and mindful cleaning~\cite{hanley_washing_2015} on well-being. Building upon research about the effectiveness of using metaphors in psycho-therapy~\cite{kopp_metaphoric_1998}, coupled with the improvements in creative performance demonstrated by the integration of embodied metaphors in VR~\cite{wang_break_2018}, each interaction was implemented using attribute or space metaphors that analogize feelings and emotions~\cite{hurtienne_sad_2009, hurtienne_soft_2016}. 
Furthermore, we incorporate the design concepts identified by \citet{wagener_letting_2023}, as well as research that established the positive impact of VR on self-reflection~\cite{wagener_selvreflect_2023}, emotions and well-being~\cite{wagener_mood_2022}, and the hands-on approach on dismantling text messages that evoke negative feelings~\cite{grieger_trash_2021}.

To evaluate \textsc{Mind~Mansion}, we conducted an explorative user study (n~=~30) where participants were confronted with 30 pre-chosen negative thoughts and engaged with them through metaphorical interactions. Our results suggest users of \textsc{Mind~Mansion} understood and enjoyed the metaphorical interactions and that it helped them to sort and remove negative thoughts, fostering a positive mindset. Furthermore, our analysis indicates a trend towards improved emotional awareness.
Based on our findings, we discuss relevant design considerations for creating a \ac{VR} experience to interact with negative thoughts metaphorically. 

This paper contributes the following: (1) the design and implementation of \textsc{Mind~Mansion} a virtual environment to support engagement with thoughts using cleaning metaphors; (2) a mixed-method user study to evaluate \textsc{Mind~Mansion}; and (3) design recommendations for \ac{VR} applications that aim to support engagement with negative thoughts as a self-care practice, informing future work in a more clinical context.

%% file: content/02_relatedwork.tex
\section{Related Work}
\label{sec:relatedwork}


This section provides an overview of traditional therapy approaches, \ac{VR}-based interventions, and the role of metaphors within the contexts of therapy, interaction design, and emotions.

\subsection{Traditional Therapy Approaches}

\ac{CBT} employs various techniques, such as questioning the validity of negative thoughts and developing alternative, realistic thoughts that better reflect one’s situation~\cite{larsson_using_2016}. \ac{ACT} creates a context in which individuals can become more self-aware of their thoughts and emotions, allowing them to develop a more accepting and compassionate attitude towards them~\cite{larsson_using_2016}. The main difference between \ac{CBT} and \ac{ACT} lies in the different meaning of cognition, as it is viewed as “a thought process in \ac{CBT} and a private behavior in \ac{ACT} ”~\cite{hofmann_acceptance_2008}. Cognitive restructuring, cognitive defusion, and emotional regulation are key features of these therapy approaches~\cite{larsson_using_2016, gross_emotion_2015}, which will be explained in the following.

\textbf{Cognitive Restructuring:}
Cognitive restructuring aims to recognize negative thoughts and examine their accuracy to take charge of them by substituting them with rational alternatives~\cite{larsson_using_2016}. This technique can also be employed to handle negative emotions~\cite{garnefski_relationship_2002}. Essentially, cognitive restructuring aims to provide a realistic perspective of the actual world rather than teaching patients positive thinking techniques or making their thoughts seem absurd~\cite{hofmann_acceptance_2008}. 

\textbf{Cognitive Defusion:}
In contrast to cognitive restructuring, cognitive defusion focuses on altering the patient’s relationship to their thoughts rather than controlling them or changing their appearance and recurrence. In other words, the approach involves observing the thoughts without attempting to counter or change them~\cite{larsson_using_2016}. The most common technique is inspired by Titchener’s “word repetition technique”~\cite{titchener_text-book_1914}, which involves repeating a word frequently until it no longer seems credible. 

\textbf{Emotion Regulation:}
Managing negative emotions is called Emotion Regulation (ER) ~\cite{gross_emerging_1998, gross_individual_2003, mcrae_emotion_2020}. ER encompasses all (un-)conscious processes that affect one's emotional responses in order to achieve a more appropriate mode of functioning and well-being~\cite{bosse_learning_2013, gross_emotion_2015}. The core of ER is to alter emotions in their intensity, quality, and duration, mainly through paying attention to specific aspects of a situation, reframing a troubling situation, or regulating the emotional reactions one has towards troubling stimuli~\cite{webb_dealing_2012}. 

\subsection{VR-Based Interventions for Emotion Regulation and Mental Well-Being}
In regard to engaging with negative emotions, VR has recently been explored as a tool to be administered to patients during advanced stages of art therapy~\cite{hacmun_principles_2018, hacmun_artistic_2021}.
However, only a limited number of studies have explored the design space of VR applications that are explicitly tailored to individuals seeking digital support for processing negative emotions in a self-care context. We argue that one of the reasons for the limited amount of research found in this area is the need to carefully design such interventions, including theoretical grounding and state-of-the-art interventions from psychology~\cite{slovak_designing_2023}. Otherwise, they can potentially re-introduce trauma and lead to rumination~\cite{wagener_role_2021}. Further, as this is a sensitive and very personal topic, personalized VR content is of utmost importance~\cite{baghaei_time_2020}.
Another illustrative case of personalized emotional content are self-created emotional islands to visualize valence in a multi-user VR setup~\cite{semsioglu_isles_2021}.

In terms of mental well-being, \citet{wagener_mood_2022} discovered that enabling users to independently create a virtual environment for visualizing their emotions is a valuable tool for emotional engagement. The results suggest that autonomous creation in \ac{VR} positively impacts emotions and well-being. 
To further explore its potential for personal reflection, another \ac{VR} application was developed, known as SelVReflect~\cite{wagener_selvreflect_2023}. This \ac{VR}  guided users through reflecting on their personal struggles and emotions with a voice-guided approach, providing a platform for free expression through painting. The findings indicate that the \ac{VR} experience allowed participants to gain a better understanding of their challenges and themselves, letting them approach their issues and their components from different perspectives.

Furthermore, previous work identified two general design requirements and four specific design concepts for emotion regulation in \ac{VR} through expert interviews with therapists~\cite{wagener_letting_2023}. Here, therapists recommended that \ac{VR} should engage users both mentally and physically to enhance cognitive involvement. It is also crucial to provide users with a space for individually interpreting metaphors using stereotypical objects, like a bench for relaxation, while also allowing them to create their own environments for personal needs. This approach aims to offer a personalized and individualized experience. 
Additionally, related work has provided frameworks for facilitating reflection in VR, for example, the ``RIOR'' model by \citet{jiang_beyond_2021}. It applies a theatrical perspective, suggesting techniques such as altering users' physical perspectives and incorporating personal item representations. However, the RIOR framework lacks specific design recommendations. In this study, we address the RIOR framework by providing common objects from a typical apartment that users may identify with.

Although further research on applications that touch on the processing of negative emotions is still scarce, \citet{grieger_trash_2021} developed two distinct concepts in a prototype: 'Positivity', where users transform negative messages into positive ones through interactions like changing their color and crossing out negative words, and 'Trash-it', where users physically engage with the message by crumpling, punching, and ultimately making it disappear. The prototype has shown promising results in positively shifting negative thoughts. All study participants reported feeling more relaxed following the \ac{VR}  interactions, and most of them noted a sense of self-reflection. The physical interaction with textual messages through gestures received positive feedback, as did the immersive and focusing nature of \ac{VR}  technology. The two distinct concepts produced different effects, as 'Positivity' offered a fresh perspective, while 'Trash-It' helped temper anger. However, it's worth noting that 'Trash-It' had a more polarized effect, with some participants finding it too aggressive. 

Apart from well-being, \ac{VR} has been proven to be effective in the treatment of anxiety disorders, including post-traumatic stress and phobias~\cite{baghaei_time_2020}. Through cognitive restructuring training, patients have been able to reduce their anxiety symptoms~\cite{rothbaum_controlled_2000, botella_virtual_2000, garcia-palacios_virtual_2002}. \ac{VR}  Exposure Therapy has been found to be equally effective as standard exposure therapy, making it a promising tool in mental health treatment~\cite{rothbaum_controlled_2000}. 

\subsection{Metaphors}

\subsubsection{Metaphors in Traditional Psychology}

To grasp the concept of emotional experiences, feelings, and situations, metaphors are often used as a tool of self-expression~\cite{kopp_metaphoric_1998}, especially when a specific word is missing to convey a certain meaning. This integrates the logic of words with the imagery of analogies to enhance comprehension of the intended message~\cite{langer_philosophy_2009}. \citet{kopp_metaphoric_1998} demonstrated that the use of metaphors in psychotherapy can assist patients in elaborating and exploring their emotions, which can then be transformed by finding similarities between the created imagery and their actual situations. By altering the image, patients gain new insights into the nature of the problem, and new opportunities for constructive problem-solving arise. 
Metaphors can assist patients in recognizing negative cognitive patterns that have become habitual and comfortable. By using emotional symbols to identify self-critical habits, patients can develop strategies to counteract them. Moreover, metaphors are useful for leveraging one's own knowledge and experiences to guide treatment efforts~\cite{otto_stories_2000}. 

\subsubsection{Metaphors in Technology}

Embodied interaction is a fundamental approach in HCI design that emphasizes the importance of everyday experiences in user-system interaction, whereas metaphorical references help users understand complex or abstract concepts by relating them to something familiar or tangible~\cite{kim_metaphors_2019, jung_metaphors_2017, carroll_hci_2003}. \citet{wang_break_2018} found that using embodied metaphors in a \ac{VR}  environment resulted in improved creative performance, with increased originality, fluency, flexibility, and persistence. Specifically, the metaphor "breaking the rules" was embodied as an action of breaking down walls, where the physical sense of "breaking" translated into conceptual processing. More specifically, participants had to 'break the walls' while navigating a corridor and were tasked with solving creativity-demanding problems under two conditions: 'break,' where they had to break virtual barriers to continue, and 'no-break,' where no barriers were present. Participants showed higher levels of originality and fluency in Alternative Uses Tests when confronted with the 'break' condition compared to the 'no-break' condition. 

Hurtienne~et~al.~\cite{hurtienne_how_2017, hurtienne_sad_2009, hurtienne_soft_2016} identified several image-schematic metaphors for mapping physical-to-abstract concepts in interaction design. These metaphors include attribute and space metaphors that can be used to draw analogies to feelings and emotions. In the context of the big~-~small attribute pair, it was associated with amount, power, importance, and significance. The bright~-~dark attribute pair is mostly associated with quality, morality, happiness, religiousness, and anger. Additionally, emotions and anger are suitable target domains for the warm~-~cold attribute pair. When an object is described as smooth, it typically conveys qualities of ease, predictability, and continuity. In contrast, when labeled as rough, it often signifies challenges, dangers, or interruptions in a process~\cite{hurtienne_sad_2009}. Further identified metaphors were “Happy is up~-~Sad is down”, “Powerful is up~-~Powerless is down”, “Good is up~-~Bad is down”, “The future is in front~-~The past is back”, “Novel is front~-~old-fashioned is back”, and “The present is near~-~The past is far”~\cite{hurtienne_soft_2016}. 

Effective use of metaphors is crucial in the development of \ac{VR}  as it determines the behavior of the virtual environment and the user's interaction with it. Clear and appropriate metaphors allow users to interact with the virtual environment in a comfortable and effective way, while poor or confusing metaphors can lead to disorientation and hinder user effectiveness. To ensure the effectiveness of metaphors in \ac{VR}, it is important to determine whether a specific task requires a particular metaphor or if there is room for flexibility to optimize the interaction. For instance, if the user is expected to feel like they are inside a building, the metaphor used should closely mimic the real-world environment of a building. On the other hand, for presenting additional information, less intrinsic metaphors can be used to trigger interactions. Ultimately, the choice of metaphor should align with the goals of the \ac{VR}  experience and enhance the user's overall experience~\cite{bryson_approaches_1995}.

\subsubsection{The Effect of Clutter and Cleaning on Mental Health}

Furthermore, \citet{lee_wiping_2011} found that the metaphor of washing away one's sins has evolved into a broader concept of starting fresh or clearing one's past, allowing individuals to get rid of unwanted baggage, whether it's related to their moral self-image, recent decisions, or concerns about bad luck. 

Complementing this, a study found that women who labeled their homes as stressful due to clutter and unfinished projects showed chronic stress levels associated with adverse health outcomes. They experienced increased depressive moods as the day progressed, following greater evening fatigue and difficulties in transitioning from work to personal life. On the other hand, women who described their homes as relaxing and often mentioned their gardens or outdoor areas were less stressed throughout the day and felt less sad. These positive effects on stress and mood remained consistent regardless of marital satisfaction or neuroticism~\cite{saxbe_no_2010}.
Other research has shown that the quality of housing can affect the socio-emotional development and cognitive abilities of children and adolescents. Poor housing quality has been linked to potential negative effects on their well-being. Additionally, housing quality can play a role in influencing levels of psychological distress~\cite{evans_child_2006}. 
These findings imply a potential connection between household stress and overall well-being, underscoring the significance of addressing stress factors in household management within the context of mental health. Therefore decluttering can be a helpful exercise in order to relieve stress and improve one’s mental well-being. 

Other research demonstrated that people with high levels of anxiety or stress often engage in repetitive behaviors like cleaning in order to regain a sense of control~\cite{lang_effects_2015}.
A study conducted by \citet{hanley_washing_2015} examined the practice of mindful dish-washing, involving participants to be mindful while washing dishes by being conscious about aspects like the scent of soap and the warmth of the water on their hands. The findings indicate that compared to the baseline of generic dish-washing, mindful dish-washing results in a 27\% decrease in nervousness and a 25\% improvement in inspiration.
In addition to mindfulness, the physical activity that inevitably accompanies cleaning is another factor in improving overall well-being because of its release of endorphins. Endorphins are neurochemicals created by the brain during exercise, moments of excitement, or in response to pain. They serve as natural pain relievers and play a role in promoting a positive sense of well-being~\cite{navines_interaction_2008}. Consequently, physical activity offers positive effects on symptoms of depression. Current research indicates that physical activity may serve as an effective therapeutic approach for both acute and chronic depression~\cite{dinas_effects_2011}.

%% file: content/03_concept.tex
\section{Designing Mind Mansion}
\label{sec:concept}

\input{img/RoomBeforeAfter}

We propose an approach to support mental health that leverages VR technology to engage with and manage negative thoughts. Building on research on decluttering and cleaning to positively impact mental health~\cite{saxbe_no_2010, hanley_washing_2015}, we designed a home setting displaying an apartment. Cleaning and physical activity are linked to beneficial effects on mental health~\cite{lang_effects_2015, navines_interaction_2008} and offer space for many common metaphors from everyday life activities applicable to coping with thoughts. For our VR application \textsc{Mind Mansion}, we used a simulated home environment resembling an apartment, with cleaning activities symbolizing coping mechanisms for negative thoughts. Users interact with objects representing negative thoughts, gradually transforming the environment from cluttered, dusty, and dark to tidy and bright. 
To design the individual interactions, we leveraged concepts, building on related work to cope with emotions and thoughts. 
Through physical interactions, users can diminish the perceived importance of negative thoughts or emotions by physically engaging with them. This process not only allows users to confront and manage their emotions in a tangible way but also empowers them by gaining a sense of control and strength relative to their emotions. Furthermore, engaging in physical activity can lead to various psychological benefits, including elevated mood, stress relief, improved coordination, and increased confidence~\cite{shosha_brief_2020}. After every interaction, we designed a further transition from darkness to brightness, associated with the "dark-bright" attribute pair ~\cite{hurtienne_sad_2009}, to evoke happiness during cleaning. Cleaner, more organized spaces signify a shift from chaos to clarity.
The concept also leverages the 'big-small' attribute pair associated with the perception of importance and significance~\cite{hurtienne_sad_2009} of internal thoughts and struggles by minimizing the size of some objects. Additionally, the concept of "Wrapping It Up," as discussed by ~\citet{wagener_letting_2023}, is evident in those interactions where objects are sorted and stored for future engagement rather than entirely eliminating them. Other objects instead are designed to disappear after the interaction, following the concept of "Letting It Go"~\cite{wagener_letting_2023}. Eliminating these physical manifestations of dirt allows individuals to let go of past neglect or disorder.
The gradual cleanup of the environment through the interaction with the individual objects and negative thoughts mirrors emotional progress, with walls changing from cold gray to warm orange. This alteration aims to create a warmer, more comfortable atmosphere as users progress through their negative thoughts, reinforcing a positive shift in emotions like happiness and excitement~\cite{yildirim_effects_2011} once all interactions are completed. After completing interactions, users are encouraged to reflect on positive changes. Hand controllers in VR simulate real-world hand movements for immersive interaction. As repetitive movements can contribute to coping~\cite{lang_effects_2015}, the experience includes ten interactions repeated three times each, presenting 30 negative thoughts in total, ensuring repetition and engagement.

\input{img/InteractionGrid}

\paragraph{Clean your Space, Clean your Mind}

Cleaning interactions are metaphorical representations of overcoming challenges and interruptions in a process, as noted by ~\citet{hurtienne_sad_2009}. Just as clutter symbolizes roughness and obstacles, cleaning involves addressing and removing these impediments, restoring both physical and metaphorical order. This process fosters clarity, happiness, and renewal. \textsc{Mind Mansion} incorporates two interactions to implement the metaphor of cleaning up.

As one way to interact with a negative thought, users can \textbf{mop the floor}, depicted in \autoref{fig:InteractionGrid}, 4A and 4B. Three dark puddles, each displaying a negative thought, are spread across the floor. As users use the mop to wipe a puddle, it gradually becomes more transparent until it ultimately disappears, taking the negative thought with it. To trigger this interaction, users grab the handle of the virtual mop and use a physical left-to-right or circular arm motion while standing upright and remain stationary otherwise.


Another cleaning interaction is implemented through \textbf{wiping windows}, as displayed in \autoref{fig:InteractionGrid}, 9A and 9B. This interaction lets users wipe three windows using a sponge by performing a physical wiping motion. Each window has a negative thought written across it and is obscured by dirt, preventing light from entering. As users wipe the windows with the sponge, the dirt gradually fades away, allowing sunlight to pour in and illuminating the floor and walls with expansive rays of light. Simultaneously, the negative thought disappears, and users can enjoy the view outside. Notably, this interaction also creates a sense of stress relief evoked through the opportunity to look at the nature outside and the slabs of sunlight illuminating the apartment~\cite{gentile_nature_2023}.

\paragraph{Sorting your Thoughts}
Organizing and structuring interactions in \textsc{Mind Mansion} serve as a visual representation of the cognitive process of sorting one’s thoughts. Users encounter a bookshelf with three fallen books and four plants lying in front of it, encouraging the user to \textbf{sort the bookshelf} and therefore also sorting the mind, depicted in \autoref{fig:InteractionGrid} 10A and 10B. 
The titles of these fallen books represent negative thoughts, and while the four plants do not specifically represent negative thoughts, they were implemented to ensure sufficient engagement in order to give users time to grasp the meaning of this interaction. Users are tasked with sorting the fallen pieces back into the bookshelf. When an object is grabbed, a purple transparent mesh in the object’s shape appears inside the bookshelf, providing guidance on where the object should be placed. This design element ensures that users maintain an organized shelf, preventing haphazard placement and the creation of additional clutter. As the object touches the purple indicator, it will snap into place, allowing the user to throw or place the object into the shelf. Once the books touch the shelf, the negative thoughts disappear from their title. To reach the objects on the ground, users have to bend over or kneel down. Once holding the objects, users have to extend their arms to reach the objects' target locations.  When this task is completed, users will find themselves in front of a nicely organized shelf with no clutter in front of it. 
To metaphorically discard negative thoughts and sort them out of the user's mind, \textsc{Mind Mansion} allows the user to engage in \textbf{throwing out trash bags} representing negativity as displayed in \autoref{fig:InteractionGrid} 5A and 5B.
For the trash bags interaction, users can grab three trash bags, each containing a negative thought depicting the metaphorical ballast one bears, and either throw or place them into the bin. Again, users must bend over or kneel to reach the trash bags. Following this, users either simulate the action of throwing the trash bag by physically raising their arm and swinging it forward, or they move to the bin to deposit the bags. Through this action, users create a physiological distance from their thoughts, hinting at the ‘The future is in front~-~The past is back’ analogy~\cite{hurtienne_soft_2016}, symbolizing concluding with a thought. 

\paragraph{Battle Negative Thinking}
Another common metaphor to cope with thoughts is to fight them. Through physical interactions, such as punching or boxing, users can diminish the perceived importance of negative thoughts or emotions by physically fighting and engaging with them, symbolized by the act of striking at an object or thought. 

By \textbf{punching pillows}, as shown in \autoref{fig:InteractionGrid}, 8A and 8B, users can deal with thoughts by minimizing their size and rearranging them. Here, users have to punch three oversized pillows on the couch, each displaying a negative thought in writing. By pressing the select button on the side of the controller, users can clench their virtual hands into fists. Users use proper physical punching, hitting, and slapping motions to shrink the pillows, similar to boxing a punching bag. When making contact with a pillow, the controller provides haptic feedback through vibrations, enhancing the sense of touch. With every punch, the size of the pillow decreases. Once it reaches a certain size, the negative thought disappears, and the pillow will no longer shrink. Users then have the option to arrange the pillows to their liking on the couch. 

Similar to the pillow~-~punching interaction, users can \textbf{beat carpets} as presented in \autoref{fig:InteractionGrid}, 3A and 3B, where they hit three carpets arranged on the walls of the apartment. In contrast to the hands-on approach of the previous interaction, users will use a virtual carpet bat to rid the carpets of dust by performing a physical beating motion. While this task requires less of the intense punching motion needed for the pillow interaction, the large size of the carpet causes users to stretch and bend over to reach all parts of it. A negative thought is written across each carpet. Initially, the carpets appear dark. However, with each strike of the carpet bat, the dust disappears, revealing the carpet's true pattern and colors. The negative thought disappears when all of the dust and dirt is removed. 

\paragraph{Deal with Bottled-up Emotions}
Inspired by the saying “bottling up emotions” we leverage the metaphor of closing up on emotions, not expressing or showing them, especially when they make one tense or angry. To deal with it we included two different approaches. 

\textbf{Watering plants} serves as an illustration of the design principle known as 'Tying It In'~\cite{wagener_letting_2023}, demonstrating how personal growth and transformation can result from accepting negative thoughts. Further, prior work~\cite{mostajeran_adding_2023} showed that virtual plants in a \ac{VR} scene increase the psychological well-being of participants. In this interaction, as shown in \autoref{fig:InteractionGrid} 6A and 6B, users grab a virtual watering can in order to water plant pots. When the watering can is in their grasp, three plant pots are highlighted with white outlines for easy identification. Users can approach each plant pot, each of which contains a negative thought written in the soil. To initiate the interaction, users raise their arms and tilt their wrists to pour water from the can, mimicking the motions of real-life plant watering. Once the users start watering the soil, a plant emerges from the negative thought. Additionally, this incorporates research that underscores the positive impact of indoor plants on well-being~\cite{berger_appearance_2022}.

Moreover, the action of \textbf{pouring bottles} embodies the concept of perceiving and recognizing a thought without a consequent removal or action. Users approach three water bottles filled with black liquid, each labeled with a negative thought. Users can pick up these bottles, and while grabbing, an empty water bottle to the left is highlighted, indicating it can be filled. This task requires users to extend their arms to reach each bottle and turn their arms over to pour the water. Although the task does not require vigorous physical activity, precise motor skills, and spatial awareness are required to prevent water spilling. The black liquid from the three bottles is sequentially poured into the empty one, and when a bottle is emptied, the negative thought written on the bottle vanishes. Although the empty bottle appears to be the same size as the others, users will observe it fits the contents of all three bottles. Notably, the black liquid transforms into a light yellow color when filled into the empty bottle.

\paragraph{Blowing off Steam}

Inflating negative thoughts, initially magnifying their importance, shrinking again upon release, or floating away can illustrate their exaggerated nature. The process of inflation also resembles 'blowing off steam', where the once inflated thoughts afterward diminish in relevance and are removed from the environment. 

Users can inflate three balloons, each labeled with a negative thought, using the microphone, embodying the metaphor of ‘blowing off steam’ and watching \textbf{rising balloons}. To float upwards, the balloon must reach a certain size, however, users are not limited to how large they wish to inflate the balloon beyond that point, allowing users to engage with this thought at their own pace. Once they’re ready, they can release the balloon, allowing it to ascend into the air, passing through the ceiling and into infinity. The white outline surrounding the balloons allows the user to follow their ascent, even after passing through the ceiling. If a balloon has not reached the necessary size to ascend, it descends, and users get another opportunity to inflate it further. 

Similar to the previous balloon~-~rising interaction, users approach three additional balloons on the table with negative thoughts written on them and enable users to \textbf{popping balloons}. Each balloon can be grabbed and inflated by blowing into the microphone. As the balloons expand, they reach a certain size, after which they pop, bursting into small pieces until they completely disappear. 
For both balloon interactions, users have to physically reach out for the balloons and use their respiratory control and timing to inflate them by blowing into a microphone integrated into the headset.

%% file: img/RoomBeforeAfter.tex
\begin{figure*}[tb!]

    \begin{minipage}[t]{.47\linewidth}
        \centering
        \includegraphics[width=\linewidth]{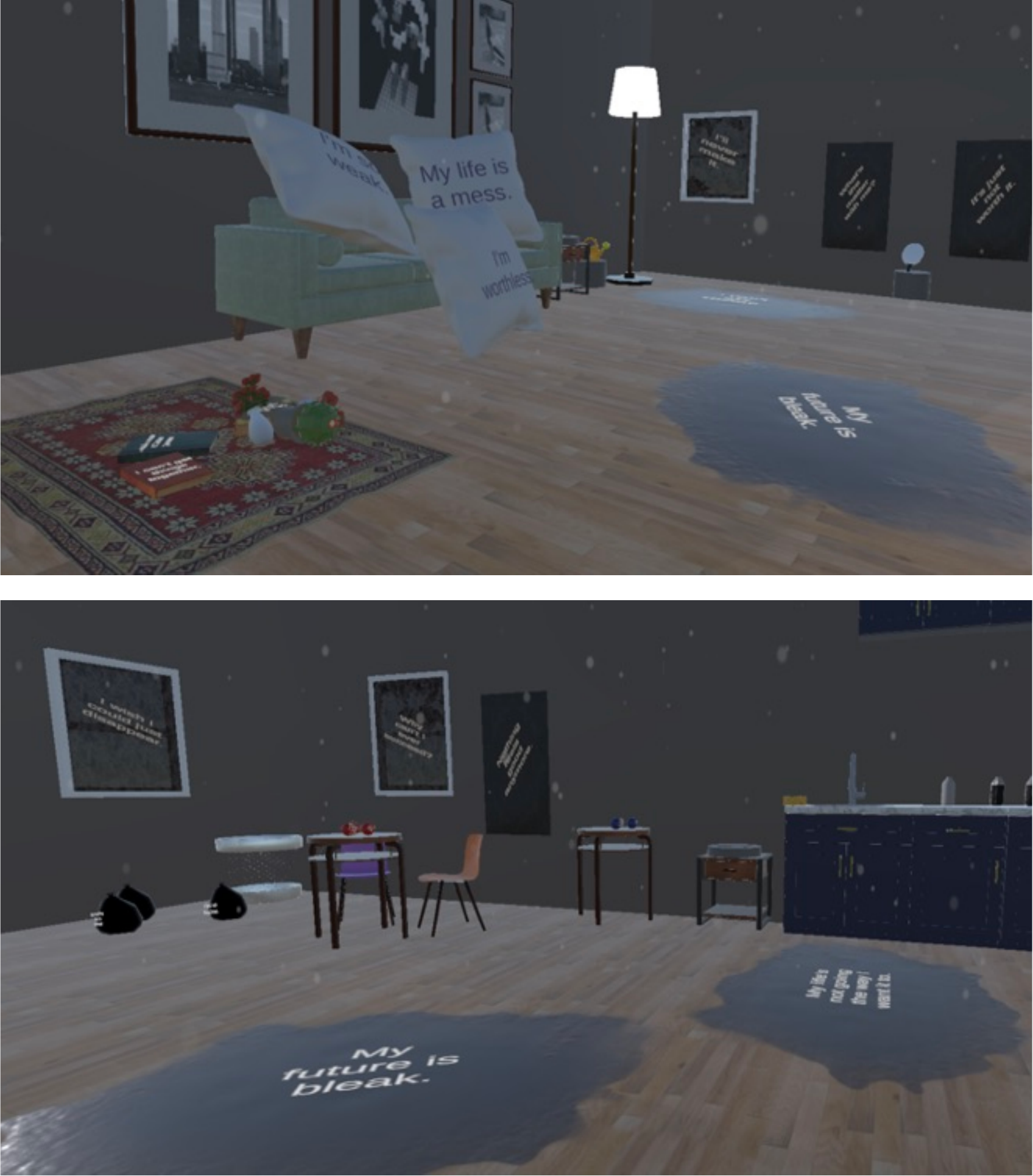}
		\subcaption{Before Interaction}\label{fig:RoomBefore}
	\end{minipage}%
    \hfill
    \begin{minipage}[t]{.47\linewidth}
        \centering
        \includegraphics[width=\linewidth]{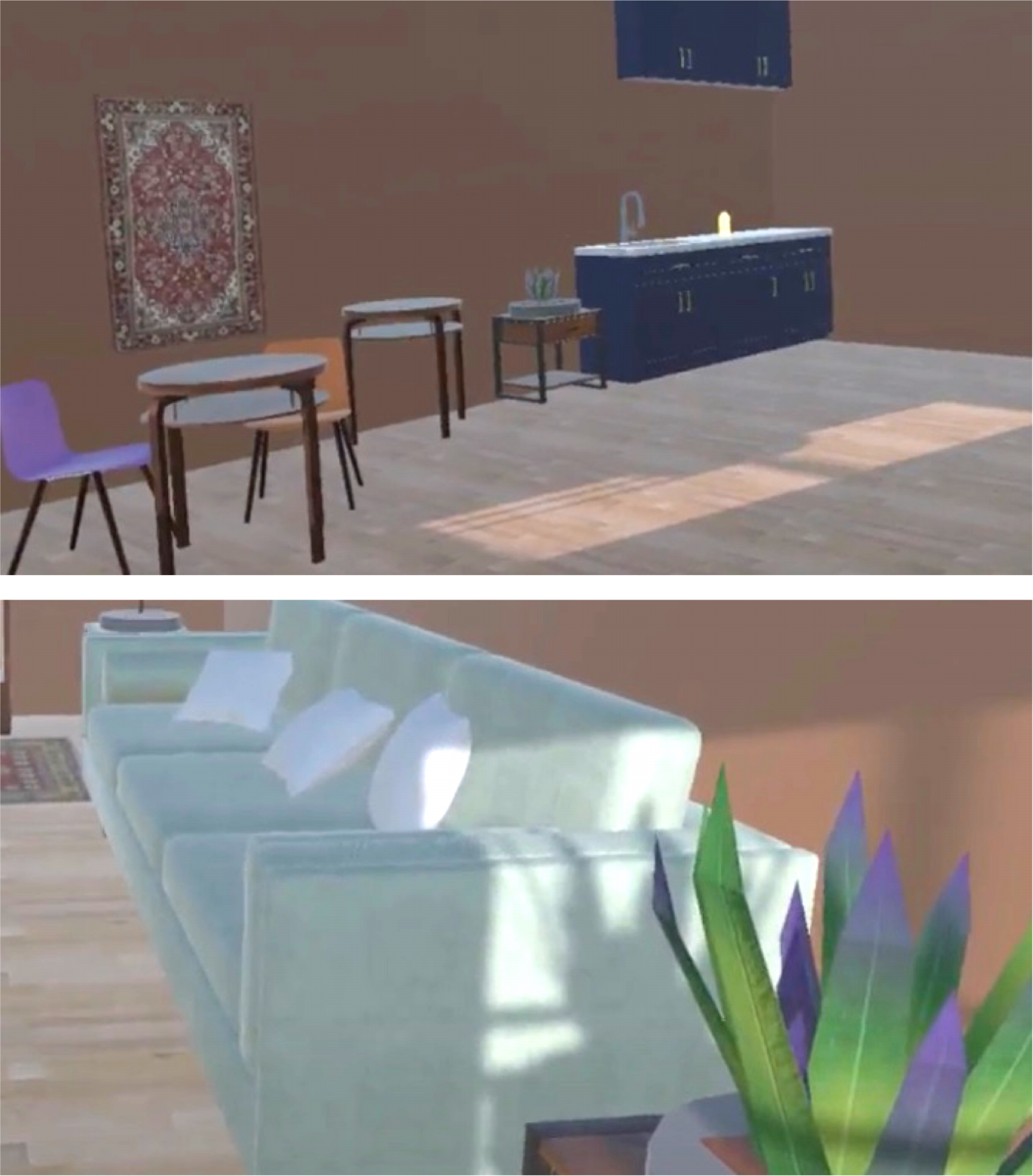} 
		\subcaption{After Interaction}\label{fig:RoomAfter}
	\end{minipage}%

 \caption{The Room of \textsc{Mind Mansion} (a) before interacting with the thoughts and (b) after interacting. Each completed interaction also increases the brightness of the room, resulting in a friendly and sorted space after finishing all interactions.}

    \Description[The room of Mind Mansion (a left) before interacting with the thoughts and (b right) after interacting]{This figure consists of four images and shows the room of Mind Mansion (a left) before interacting with the thoughts and (b right) after interacting. The left two images show the virtual apartment before the users' interactions. Here the room appears dark and dirty. Text in the form of negative thoughts is spread across the room. The right two images show the same virtual apartment after the interactions. All text is now gone, together with the mess from before. The room now appears bright and friendly.}
	\label{fig:RoomBeforeAfter}
 
\end{figure*}

%% file: img/InteractionGrid.tex
\begin{figure*}[tb!]
	\includegraphics[width=\linewidth]{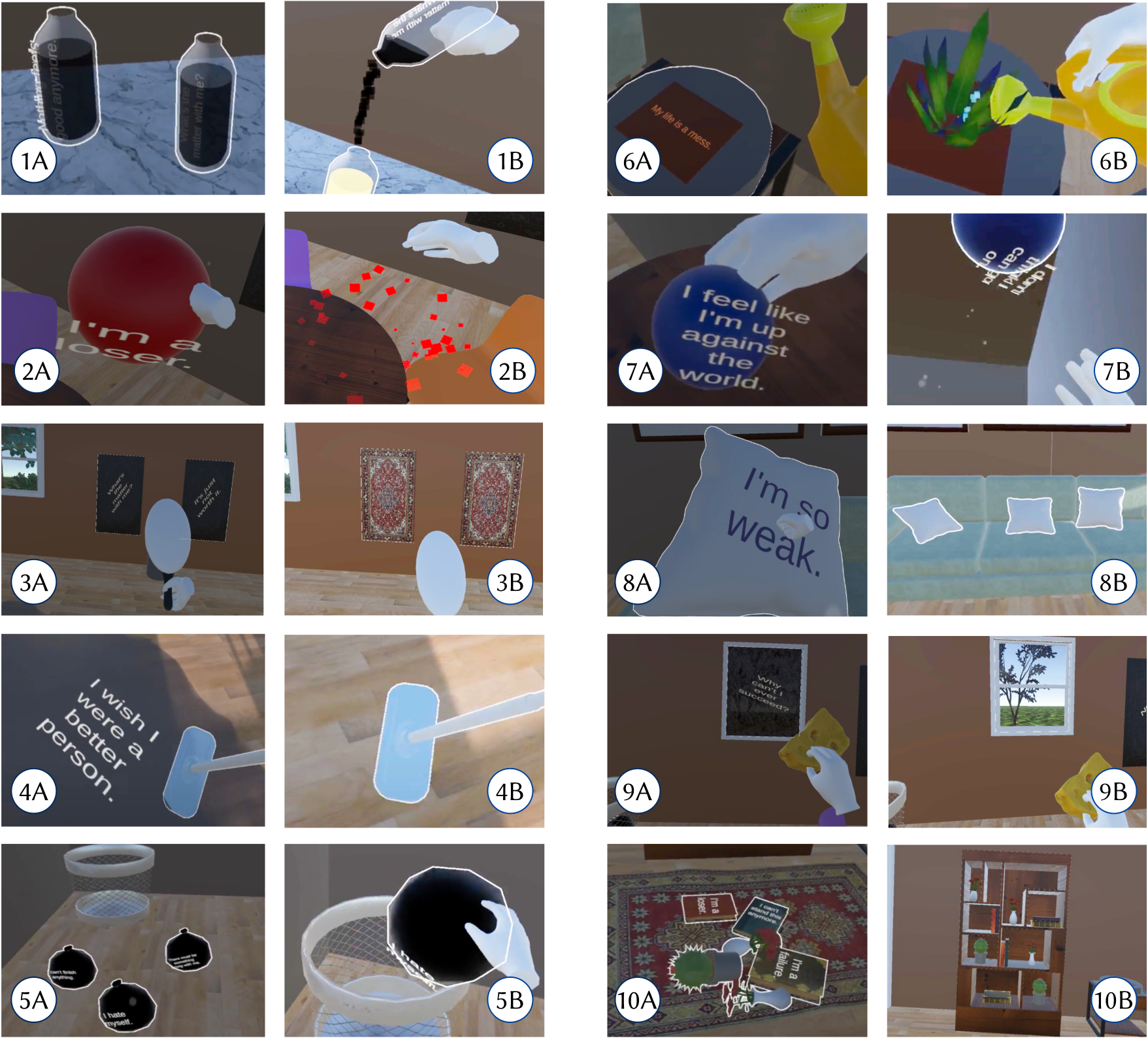}
    \centering
    \caption{An overview of the 10 different \ivMetaphoricalInteraction{}s each (xA) before and (xB) after the interaction.
    1 \ivBottleRefilling{}, 
    2 \ivBurstingBalloons{},
    3 \ivCarpetBeating{},
    4 \ivPuddleSweeping{},
    5 \ivTrashbagThrowing{},
    6 \ivWateringPlants{},
    7 \ivRisingBalloons{},
    8 \ivPillowPunching{},
    9 \ivWindowWiping{}, and 10 \ivBookshelfSorting. }

    \Description[An overview of the 10 different metaphorical interactions each (xA) before and (xB) after the interaction]{An overview of the 10 different metaphorical interactions each (xA) before and (xB) after the interaction. A grid of a total 20 images showing the Bottle Refilling, Bursting Balloons, Carpet Beating, Puddle Sweeping, Trashbag Throwing, Watering Plants, Rising Balloons, Pillow Punching, Window Wiping, and Bookshelf Sorting interaction.}
	\label{fig:InteractionGrid}
 
\end{figure*}

%% file: content/04_methodology.tex
\section{Methodology}
\label{sec:methodology}

For the evaluation of this prototype, we conducted an exploratory user study involving 30 participants, following related works (e.g., \cite{grieger_trash_2021, semsioglu_isles_2021, wagener_selvreflect_2023}). Other conventional ER interventions could integrate confounding factors. For instance, real-life cleaning cannot offer the metaphoric changes of the intervention, mobile phone applications for emotion regulation (e.g., \cite{rozgonjuk_emotion_2021, shi_instant_2023}) limit autonomy in 3D spaces, and similar VR approaches in that regard, such as by \citet{grieger_trash_2021} do not provide metaphoric interactions linked with decluttering. Thus, as our aim is to provide an initial exploration of how metaphoric interactions in VR can support emotional awareness of negative thoughts, we decided on an exploratory study design. 
The goal of this study was to evaluate the impact of \textsc{Mind~Mansion} on users, focusing on emotional states and awareness of thoughts. Additionally, we studied participants’ perceptions of negative thoughts in the context of different metaphors and interaction styles and how this affects the ability to cope and engage with these thoughts. We counterbalanced the set of 30 negative thoughts and the 10 interactions using a Balanced Latin Square design. We designed and conducted the user study in compliance with our institution's ethics committee guidelines.

\input{img/LikertPlotInVRQs}

\subsection{Data Collection and Evaluation}
We chose a mixed-method approach by combining quantitative data gathered from 
questionnaires and qualitative data through semi-structured interviews. This mixed-method approach enhances the breadth and depth of understanding and strengthens the validity and reliability of the findings by triangulating data~\cite{greene_toward_1989}.

For the quantitative data, we collected answers from each participant, using a custom in-VR questionnaire after each condition as well as a custom post-VR questionnaire after completing all conditions. 

As a qualitative method, we conducted semi-structured interviews at the end of our user study with an average time of five minutes. During the interview, participants were asked about their overall impressions of the experiment, their preferred and least preferred interactions, the reason behind their preferences, possible mood shifts, the takeaways from the experience, and their expectations regarding encountering negative thoughts in the future.
The interview's audio recordings were transcribed verbatim and analyzed in ATLAS.ti. 
We used thematic analysis to systematically examine the data~\cite{blandford_qualitative_2016}. Initially, one researcher coded 10\% of the interviews, and subsequently, three researchers discussed and agreed on the final codes. Following this, one researcher coded the remaining interviews accordingly. After this, two researchers deliberated on how to organize and group the interview codes, which resulted in three overarching themes presented in \autoref{sec:results}.

\subsection{Participants}
We recruited n = 30 participants (male:12, female:18, diverse:0)  via social networks and our institution's email distributor for user studies, with an average age of 26 years ranging from 22 to 62 years. Among the participants, 12 had no prior experience with \ac{VR}, 13 were sporadic users, and 5 identified themselves as experienced \ac{VR} users. We compensated participants with an equivalent of 10\,€. All participants were required to confirm their current mental stability and health. 


\subsection{Apparatus}
The application was developed using Unity version 2021.3.16f and operated on an HTC Vive Pro. The same \ac{HMD} and corresponding controllers were used throughout this study.
The negative thoughts featured in the \ac{VR}  experience were drawn from the \ac{ATQ}~\cite{steven_d_automatic_2011}, consisting of a total of 30 common negative thoughts that pop into people’s minds, in order to keep the study comparable.

\subsection{Study Procedure}
After welcoming the participants and introducing them to the study’s topic and procedure, participants were asked to fill out a consent form, confirm they were without a history of mental illnesses, and complete a demographic survey. Following this, participants were presented with the \ac{ATQ} to get familiar with the negative thoughts by reading and rating the negative thoughts on the accompanying scale. 
Following the questionnaires, the experimenter explained the controls and let participants familiarize themselves with the system and each of the \ivMetaphoricalInteraction{}s in a tutorial scene. Here, no negative thoughts were displayed yet. After clarifying potential questions regarding the controls, the experimenter started the \textsc{Mind~Mansion} application. Before actively engaging with the application, participants were advised to take in the virtual apartment and reflect on the emotions it evoked. Once participants felt ready to proceed, condition 1 of the experiment started. The experiment consisted of 10 conditions, each corresponding to an interaction described in \autoref{sec:concept}. Both the set of 30 negative thoughts and the 10 interactions were counterbalanced using a Balanced Latin Square design. 
As described in more detail in \autoref{sec:concept} for each condition, participants had to repeat each \ivMetaphoricalInteraction{} 3 times to ensure repetition.
After completing a condition, participants filled out the in-VR questionnaire that appeared once the interaction was finished and then progressed to the next condition. We chose this in-VR approach to let participants perceive the gradual changes of the virtual apartment.
Upon completing all 10 conditions, participants were again encouraged to look around the room and reflect on their emotions. When they felt ready, participants took off the HMD and returned to the computer to fill out the post-VR questionnaire. 
To conclude the study, we conducted a semi-structured interview, and participants received their reimbursement.

%% file: img/LikertPlotInVRQs.tex
\begin{figure*}[ht!]
	
	\centering
	\includegraphics[width=\textwidth]{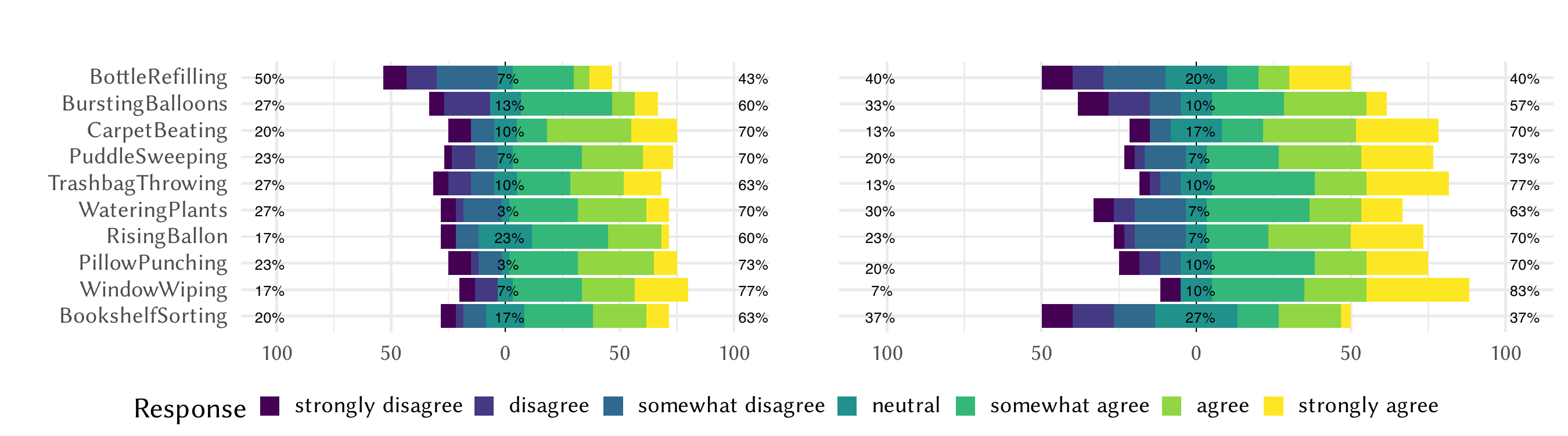}\hfill
	\vspace{-1em}
	\begin{minipage}[t]{.5\linewidth}
		\centering
		\subcaption{\dvEngagement{}}\label{fig:results:engagement}
	\end{minipage}%
	\begin{minipage}[t]{.35\linewidth}
		\centering
		\subcaption{\dvCoping{}}\label{fig:results:coping}
	\end{minipage}%
	
	\caption{Participants' in-VR responses regarding (a) \dvEngagement{} and (b) \dvCoping{} with a thought through the 10 \ivMetaphoricalInteraction{}s on a 7-point Likert scale.}
    \Description[Two 7-point Likert plots showing the participants responses to "Engagement" and "Coping"]{Two 7-point Likert plots showing the participants responses to the in-VR questionnaire regarding "Engagement" and "Coping" with the presented thoughts through the different metaphorical interactions.}
	\label{fig:results:inVRQs}
\end{figure*}

%% file: content/05_results.tex
\section{Results}
\label{sec:results}
In this section, we report the results of our user study. We report the quantitative results from our in-VR questionnaire and post-VR questionnaire before presenting the qualitative results.

\subsection{Quantitative Results}
We analyzed the in-VR Likert questionnaire as non-parametric data using an \ac{ART} ANOVA suggested by \citet{wobbrock_aligned_2011}. We followed the ART-C procedure \cite{elkin_aligned_2021} for post-hoc testing of the significant results.
Here, we report the generalized ETA squared \ges{} as a measure of the effect and classify it according to \citet{bakeman_recommended_2005}. For effect size classification, we follow the suggestions by \citet{cohen_statistical_2013} for small ($>.0099$), medium ($>.0588$), or large ($>.1379$) effect size. In this subsection, we present the results of our statistical analysis.

\subsubsection{\dvEngagement{} "Through This Interaction I Felt Engaged With the Thought"}
We found a significant (\ano{1}{9}{3.46}{<.001}) main effect for the \ivMetaphoricalInteraction{} on the \dvEngagement{} rating with a \efETAsquared{0.10} effect size. Post-hoc tests revealed significantly lower ratings for \ivBottleRefilling{} compared to \ivCarpetBeating{} ($p<.001$) and \ivWindowWiping{}  ($p<.00$). The results are visualized in \autoref{fig:results:engagement}.

\subsubsection{\dvCoping{} "This Interaction Helped Me Cope With the Thought"}

We found a significant (\ano{1}{9}{4.84}{<.001}) main effect for the \ivMetaphoricalInteraction{} on the \dvCoping{} rating with a \efETAsquared{0.14} effect size. Post-hoc tests revealed significantly lower ratings for \ivBookshelfSorting{} compared to \ivCarpetBeating{} ($p<.01$), \ivTrashbagThrowing{}  ($p<.01$) and \ivPuddleSweeping{}  ($p<.05$). We also found significantly higher ratings for \ivWindowWiping{} compared to \ivBookshelfSorting{}  ($p<.001$), \ivBottleRefilling{}  ($p<.01$) and \ivBurstingBalloons{}  ($p<.05$).  The results are visualized in \autoref{fig:results:coping}.

\subsubsection{Desired Future Usage "If I Were to Use This Application on a Regular Basis I Would Like to Use the Following Interactions"}
Looking at the results of our post-study questionnaire regarding desired future usage, we can see differences between the interactions. \autoref{fig:AfterVRQNumberRepeat} shows how many participants indicated a desired future usage for each of the 10 \ivMetaphoricalInteraction{}s. In a hypothetical regular future usage, participants mostly want to interact with \ivWateringPlants{} (22), \ivCarpetBeating{} (22), \ivWindowWiping{} (20), and \ivPillowPunching{} (20) again. Also, the \ivRisingBalloons{} (18) and \ivTrashbagThrowing{} (17) interaction was mentioned by more than half of the 30 participants. Less mentioned for future usage were the \ivPuddleSweeping{} (10), \ivBookshelfSorting{} (9), and \ivBurstingBalloons{} (8) interactions. The least mentioned was the \ivBottleRefilling{} (5) interaction, which still 5 participants would like to use in a future application. Consequently, none of the 10 interactions was unfavorable in a way that no participant would like to use them again.

\subsubsection{Impact on Daily Life "Which of These Interactions Do You Think Will Have an Impact on Your Negative Thoughts in Your Daily (Non-VR) Life?"}
Also, in the results of our post-study questionnaire regarding the impact on daily life, we can see differences between the 10 \ivMetaphoricalInteraction{}s in \autoref{fig:AfterVRQNumberImpact}. When asked about what interaction participants think will have an impact on their daily life, most mentioned the \ivWateringPlants{} (20) interaction and \ivTrashbagThrowing{} (17). Half of all participants also mentioned the \ivBookshelfSorting{} (15) and \ivWindowWiping{} (15). Less often, participants mentioned the \ivCarpetBeating{} (14), \ivPillowPunching{}, and \ivRisingBalloons{} (10). Least mentioned was the \ivPuddleSweeping{} (8), \ivBurstingBalloons{} (5), and \ivBottleRefilling{} (5). Again, with 5 mentions at least no interaction was perceived impact-less by everyone looking at our group of 30 participants.

\input{img/AfterVRQs}

\subsection{Qualitative Findings}

Based on the qualitative results, three themes were derived from the interviews: \textit{Engagement}, \textit{Externalizing and "Tangibility"}, and \textit{Impact Beyond VR}. The findings are specified below and depicted with quotes from the interviews.

\subsubsection{Engagement}
The first theme explores how \textsc{Mind~Mansion} motivated participants to engage with negative thoughts and how it increased the awareness of the effects that negative thoughts can have. In that regard, participants understood that \textsc{Mind~Mansion} provides \textit{"a personalized way with an overview of my own feelings and how things make me feel right now" (P25)}. Through Mind Mansion, they acknowledged the need to deal with negative thoughts for their well-being, as one participant described as follows: \textit{"It felt like a releasing journey, and [...] it really helped with negative thoughts" (P16)}. Participants further reflected that it \textit{"would be helpful for people to feel more engaged with the thoughts" (P16)} and that \textsc{Mind~Mansion} is seen as an application that encourages such an engagement: \textit{"[through Mind Mansion] it is good to be able to actually get in touch with the thoughts" (P14)}.
In that regard, participants also emphasized that they enjoyed physically engaging with their thoughts by repeatedly interacting with them. They felt it helpful to \textit{"actually doing something"} (P26) and to feel empowered through the physical interactions, \textit{"because I could do it myself"} (P14).

\subsubsection{Externalizing and "Tangibility"}
Besides getting motivated to deal with negative thoughts, participants also reflected on the main reasons why \textsc{Mind Mansion} encouraged engagement. Thus, the second theme revolves around the benefits of externalizing and making thoughts "tangible" in VR. In that regard, participants felt that externalizing negative thoughts and making them visible in a space around them was therapeutic and helpful. One participant stressed: \textit{"I think for the people who don’t have a good way of visualizing things in their mind, it’s really helpful" (P14)} and another one reflected that \textit{"giving your thoughts a face or a form of interaction is more powerful than you think [because] it can get quite confusing in your own head, so giving them a space in the external world or digital world is quite powerful" (P28)}. They also specified that externalizing negative thoughts facilitated the process of dealing with negative thoughts, as it provides step-by-step instructions, and the interactions can be seen as a progress bar motivating users to keep on dealing with their thoughts. P25 described this aspect as follows: “\textit{It made me feel in control. Seeing it and having it all around me or in front of me felt like, 'okay, this is what I have to deal with', and then I can do it step by step and have a clearer image of 'How far am I? Is there a progress?'}” (P25).

Further, participants identified that the "tangibility" of negative thoughts in VR, alas not a haptic experience, helped them to process and manage negative thoughts. One participant elaborated: \textit{"your thoughts [become] a concrete image that you can work with or touch" (P28)}. This "tangibility" led to participants feeling empowered to deal with the negative thoughts, and through interacting with them, they felt in control. One participant explained: \textit{"I could hold them [the negative thoughts], and then it was my control to let them go" (P10)}.

\subsubsection{Impact Beyond VR}

The third theme encompasses comprehensive insights into the lasting effects of \textsc{Mind~Mansion} beyond the scope of the VR experience, reflecting on and learning new skills usable in real life. 
At first, participants reported that through \textsc{Mind~Mansion}, they started to reflect on their strategies used in real life to cope with negative thoughts. They thought that if they would \textit{"play this game, then maybe it helps me with my negative thoughts and getting a new perspective" (P16)}. Further, they emphasized reflecting on how negative thoughts affect them in their daily lives, perceiving \textsc{Mind~Mansion} as \textit{"a metaphor how negative thoughts affect my view and my acting in life" (P23)}. They also started thinking about strategies when dealing with negative thoughts in reality. To elaborate, some participants emphasized the importance of accepting negative thoughts. For instance, P25 said: \textit{"it made me realize that these low phases are normal or even important, you just take a moment and confront them, but still knowing you can move on and just create something new and beautiful for yourself"} (P25) while other participants learned about the importance of actively managing negative thoughts: \textit{"it is a quiet corner to think about how this might affect me in the future, because I feel like holding back these thoughts can have a negative impact on life}” (P31)

Additionally, participants reported that the interactions felt familiar because they act similarly in real life when experiencing negative thoughts. To say this with the words of P8: \textit{"[The pillow punching] is an action that I think most of us have done before. When you feel sad, you punch something. And pillow punching is very close to what most people have punched before.” (P8)}
In turn, participants were also inspired by the VR interactions to use them in reality to cope with negative thoughts. One participant explained: \textit{"this virtual reality experience can start to get [into] an interaction routine, it can copy to real life"} (P12).

%% file: img/AfterVRQs.tex
\begin{figure*}[tb!]
	\begin{minipage}[t]{.5\linewidth}
		\includegraphics[width=\linewidth]{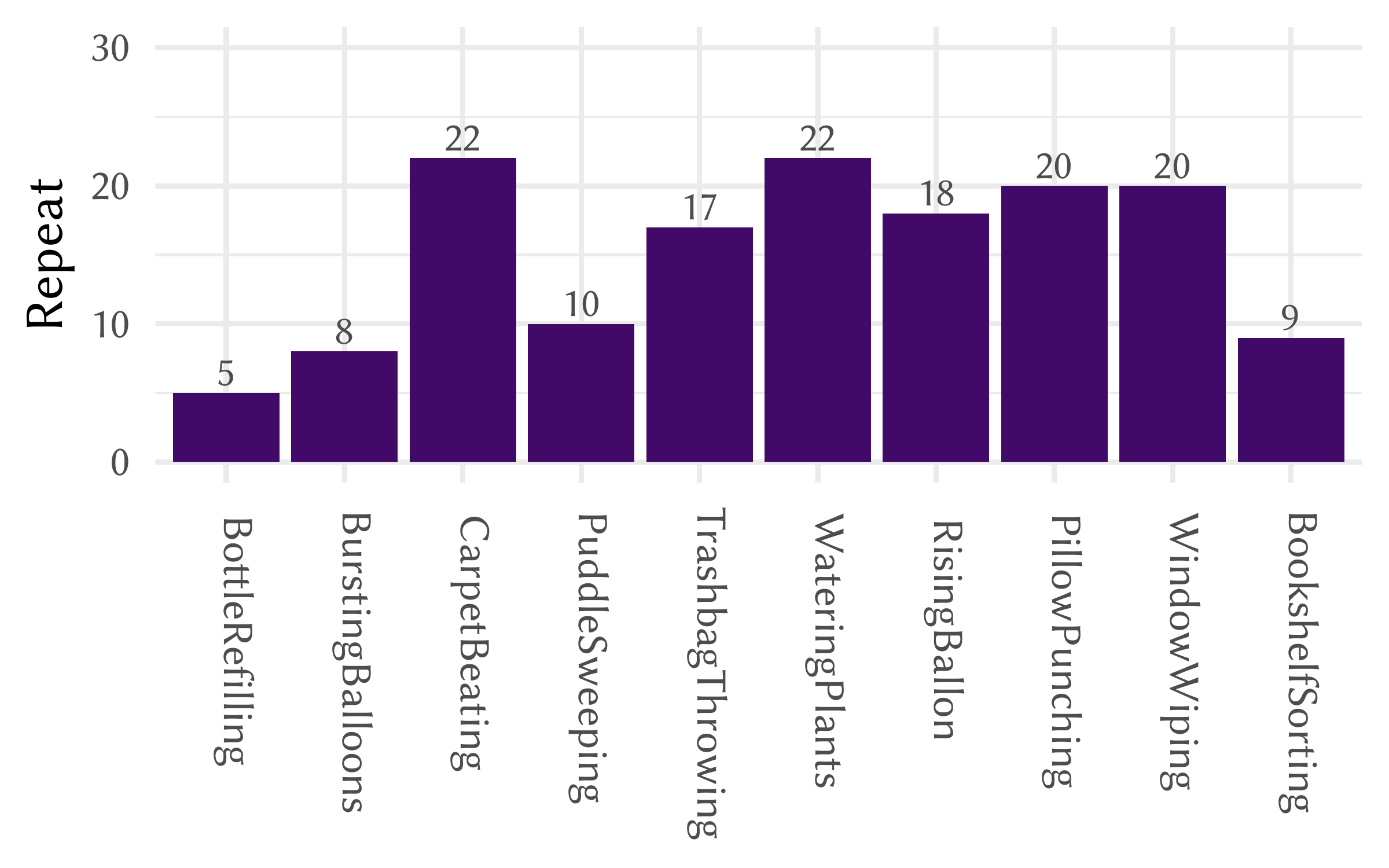}
        \centering
		\subcaption{Desired Future Usage}
        \label{fig:AfterVRQNumberRepeat}
	\end{minipage}%
    \begin{minipage}[t]{.5\linewidth}
		\includegraphics[width=\linewidth]{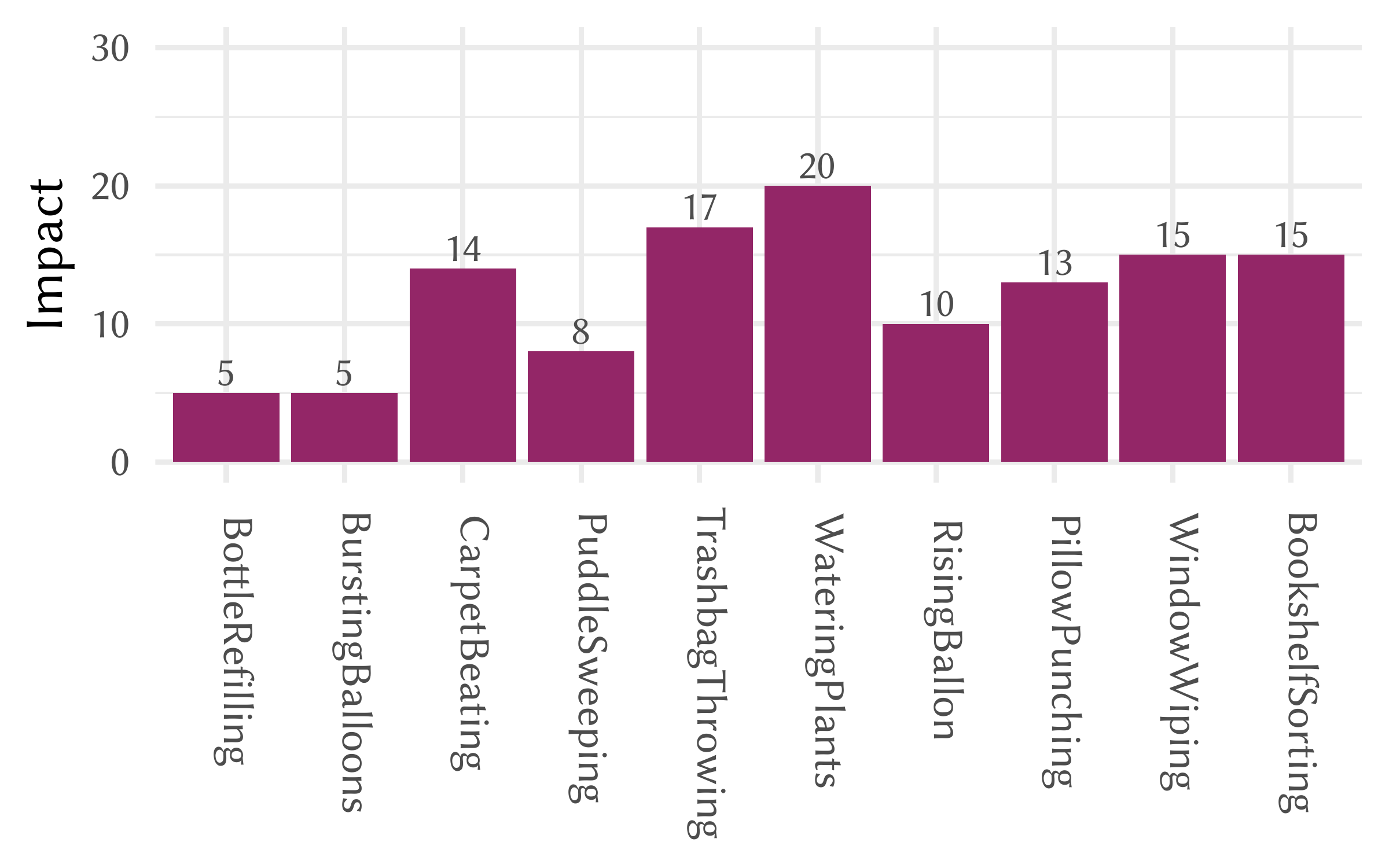}
        \centering
		\subcaption{Impact on Daily Life}
        \label{fig:AfterVRQNumberImpact}
	\end{minipage}%

 \caption{The results of the post-VR questionnaire visualized as a histogram showing how often each interaction was mentioned by participants for (a) desired future usage and (b) impact on daily life.}

    \Description[The results of the post-VR questionnaire visualized as a histogram showing how often each interaction was mentioned by participants, for (a) desired future usage and (b) impact on daily life.]{This Figure shows the results of the post-VR questionnaire visualized as a histogram showing how often each interaction was mentioned by participants, for (a left) desired future usage and (b right) impact on the daily life. Each histogram has one bar for each of the 10 interactions.}
	\label{fig:AfterVRQ}
 
\end{figure*}

%% file: content/06_discussion.tex
\section{Discussion}
\label{sec:discussion}


Based on the quantitative and qualitative findings of our user study we derive the following design recommendations for effectively engaging with negative thoughts in \ac{VR}: \textit{Design for Physical and Mental Engagement} with a balance of challenge and effort in the task, \textit{Give Thoughts a Body} to foster awareness and externalization, \textit{Make the Positive Progress Visible}, and \textit{Use Familiar Interactions to allow Users to "Take Something Back to Reality"}.


\subsection{Design for Physical and Mental Engagement}
Aligning with the concepts identified by \citet{wagener_letting_2023}, the user study demonstrated that both mentally and physically engaging interactions positively influence the user's ability to cope with the corresponding thoughts. The findings also emphasize the importance of providing sufficient haptic feedback to enhance immersion and the user's connection to their mindset. In contrast to the findings of \citet{grieger_trash_2021}, interactions that were more physically engaging and could potentially be perceived as ‘aggressive’, such as punching and beating, did not lead to polarized reactions among participants. Instead, they were generally perceived as very helpful and likable. One possible explanation for this contrast could be the absence of interpersonal elements in the negative thoughts of this experience, unlike the negative text messages used in the ‘Trash-It’ experiment. This absence of personal connections to other people may have made it easier for individuals to engage in more physically aggressive interactions. Additionally, interactions should strike a balance: they should be challenging enough to engage the user to reflect on the thought without causing frustration or feelings of failure while not requiring excessive effort that might reduce the overall enjoyment and motivation of the experience, as emphasized by \citet{agapie_longitudinal_2022}.

Moreover, providing users with a sense of control without stronger limitations has proven to be an important consideration for this interaction design. Allowing users their own time to engage with each thought has helped them track their progress in coping with the associated thoughts and has empowered them to feel more confident approaching the thoughts. In contrast, interactions that were more challenging or proved difficult to handle diminished users’ confidence, making them not feel in control anymore.

\subsection{Give Thoughts a Body to Foster Awareness and Externalization}
This allows users to "face" a thought and have a representation to relate to. Our findings show that users appreciated a "tangible" representation of the thoughts, which is in line with previous works \cite{brinol_treating_2013, brinol_objectification_2017}.
By externalizing the thought, users could take a different perspective and question it. Also, when not interacting with a thought, users perceive the representation and can accept its existence.

\subsection{Make the Positive Progress Visible}
Our \ivMetaphoricalInteraction{}s benefited from clear visual representations of the progress. Visual outcomes that created a more positive environment, such as filling the room with sunlight, allowing users to see nature, and growing plants, helped users recognize they have created something positive while acknowledging and working on the negative thoughts, aligning with the “Tying~It~In” concept~\cite{wagener_letting_2023}. Furthermore, the ability to visibly track progress during interactions, such as observing the pillow shrinking with each punch or the carpet lightening with each beat, helped users to feel more engaged with the thoughts and improved their ability to realize the progress.

\subsection{Use Familiar Interactions to Allow Users to "Take Something Back to Reality"}
The use of appropriate and clear metaphors has demonstrated varying effects on how effectively users cope with negative thoughts and understand the coping process~\cite{bryson_approaches_1995}. 
When users embodied certain metaphors, two distinct associations emerged where some users felt they created something positive, while others perceived an enhancement of something negative. In contrast, some metaphors remained incomprehensible and, therefore, had no impact on how participants coped with the associated thoughts. This can be prevented by using more stereotypical objects and allowing users to create their own space, as suggested by \citet{wagener_selvreflect_2023}. However, the effectiveness of certain metaphors could be improved with additional guidance on the underlying purpose of the interaction. For instance, while people technically understood the concept of “Wrapping~It~Up”~\cite{wagener_letting_2023}, some participants were not open to allowing themselves to store a thought and leave it for future exploration, preferring instead to remove it entirely.

Therefore, interactions and metaphors that participants found familiar proved to have a positive impact on their ability to cope with the thoughts. Connecting interactions to real-life coping mechanisms, such as cleaning in general~\cite{lang_effects_2015} or using boxing as a stress relief~\cite{shosha_brief_2020}, for instance, punching pillows, helped users quickly grasp the underlying meaning behind these interactions and coping more effectively with the thoughts. 

Additionally, using these familiar interactions allows users to connect the in \ac{VR} experience with an interaction in the physical world. We speculate that this allows for a more persistent effect of the engagement, as the physical repetition could be linked to the \ac{VR} experience. 

%% file: content/07_limitations.tex
\section{Limitations and Future Work}
\label{sec:limitions}
We opted for an explorative user study in a lab environment, which comes with the following limitations, requiring future studies.

\subsection{Limited Interaction Time}
In the context of our explorative study, the \ac{VR} experience lasted approximately 20-25 minutes, and each participant went through each condition only once. One participant mentioned that while the \ac{VR} provided some insights into approaching negative thoughts, the amount of thoughts and interactions presented in the experience made it challenging to reflect comprehensively within the limited duration of the study. To potentially enhance the impact on how users cope with negative thoughts in the future, future studies may consider offering users the opportunity for repetition of the \ac{VR} experience, involving fewer interactions that participants can choose for themselves. Furthermore, there was no follow-up study to assess whether the impact of \textsc{Mind~Mansion} persisted over time. Future research should include a follow-up study to monitor the long-term impact of \textsc{Mind~Mansion} on individuals' responses to negative thoughts and how repeated \ac{VR} experiences may influence real-life behavior.

\subsection{Missing Comparison to Similar Approaches}
To keep the overall study time within a reasonable time frame, we opted against a direct comparison to other existing approaches. This allowed us to explore the presented \ivMetaphoricalInteraction{}s more in detail, to gain insights for future iterations of \textsc{Mind~Mansion}.
Consequently, the application could not be assessed in comparison to a different pre-existing application, as it introduces a new approach to engaging with negative thoughts. For example, it goes beyond confronting negative messages that evoke negative feelings in VR~\cite{grieger_trash_2021} by directly addressing the thoughts themselves. Additionally, the application directly used metaphors to confront these thoughts, addressing research on metaphors associated with emotions~\cite{hurtienne_sad_2009, hurtienne_soft_2016}, and the embodiment of these metaphors, which extends research on creativity-demanding problem solving through embodied metaphors~\cite{wang_break_2018}. 

Therefore, identifying a suitable baseline for comparison with this prototype proved difficult. Due to this experience being new, it was important to first explore how it enabled users to perceive and cope with negative thoughts, as was the focus of this work. However, based on the result gathered in this experiment concerning the relationship between different design concepts and their effectiveness in coping with the associated thoughts, future studies may consider these findings as hypotheses to be further explored. 

\subsection{Technical Limitations of VR Environments}
While the application was carefully designed, issues with controls, for example trying to grab bottles and without spilling the liquid, presented challenges in immersing into the \ac{VR} and effectively confronting negative thoughts. Consequently, some users reported not experiencing any noticeable shift in their emotions during the experience. For this reason, future applications should consider using objects that require less precise motor skills, allowing a more accessible transition into the \ac{VR} for all users. 
Furthermore, because the participants were engaged in physical activity, some of them noted experiencing sweating around their eyes. This discomfort, which could cause the \ac{HMD} to slip or simply feel uncomfortable, ultimately hindered their sense of immersion by serving as a constant reminder of the \ac{VR} headset they were wearing. 

\subsection{Predefined Set of Negative Thoughts}
To ensure the comparability of this study, all participants received the same set of negative thoughts based on related work, making it harder for participants who did not relate to some thoughts to experience a shift in their emotions. Future studies could further explore the effect of similar tools when using personalized thoughts in the experiment. 

\subsection{Therapeutic Value}
While mental health professionals recognize the advantages of self-care at home, it is essential not to underestimate the potential of falling into recurring negative emotional patterns~\cite {eikey_beyond_2021}. As seen in the study, issues arose with interactions that were either too challenging or not understandable, potentially increasing negative thoughts. These interactions present a risk, especially for vulnerable users, when not guided by professionals~\cite{wagener_role_2021}. This suggests a more in-depth examination of the specific design elements that support coping and the need for research on using this application with therapeutic guidance. 
In summary, the findings of \textsc{Mind~Mansion} indicate support in engagement with negative thoughts and a decrease in negative affect. However, it is important to note that these results are specific to this particular experiment. Therefore, future research should address this subject again and involve experts for a more comprehensive exploration.

%% file: content/08_conclusion.tex
\section{Conclusion}
\label{sec:conclusion}
With our explorative study around designing \textsc{Mind~Mansion}, we gained insights into how metaphorical interaction in a \ac{VR} experience can be used to engage with otherwise abstract thoughts. \textsc{Mind~Mansion} was developed, drawing upon existing literature and building upon established concepts, and invites users to engage with thoughts in a novel way. The \ac{VR} experience immerses users in a virtual apartment, enabling them to engage with negative thoughts through the embodiment of spatial and attribute metaphors. This interaction helps users sort and remove negative thoughts by sorting and disposing of objects in a virtual apartment, ultimately fostering a more organized and positive mindset. In our user study (n~=~30), we evaluated this approach using exemplary negative thoughts derived from the \ac{ATQ}, mapped onto the objects in the virtual apartment.
\textsc{Mind~Mansion} empowered participants to draw connections between real-life challenges and the interactions integrated into this experience. This, in turn, enabled them to approach new perspectives on negative thoughts, embrace them, and identify constructive ways to deal with negative thoughts in the future. 
Our design recommendations for designing similar \ac{VR} systems are 1) \textit{Design for Physical and Mental Engagement} with a balance of challenge and effort in the task, 2) \textit{Give Thoughts a Body} to foster awareness and externalization, 3) \textit{Make the Positive Progress Visible}, and 4) \textit{Use Familiar Interactions, to allow Users to "Take Something Back to Reality"}.
In summary, \ac{VR} interventions for interacting with negative thoughts through embodied metaphors are a promising tool to confront negative thoughts and foster awareness of emotions. However, finding the right balance between allowing users some independence and effectively addressing negative thoughts, carefully implementing the embodied metaphors is crucial to prevent users from becoming trapped in recurring negative emotions. This needs further future investigation, and we see \textsc{Mind~Mansion} as an early prototype of this journey.



%% file: mindmansion.bbl

\begin{thebibliography}{67}


\ifx \showCODEN    \undefined \def \showCODEN     #1{\unskip}     \fi
\ifx \showDOI      \undefined \def \showDOI       #1{#1}\fi
\ifx \showISBNx    \undefined \def \showISBNx     #1{\unskip}     \fi
\ifx \showISBNxiii \undefined \def \showISBNxiii  #1{\unskip}     \fi
\ifx \showISSN     \undefined \def \showISSN      #1{\unskip}     \fi
\ifx \showLCCN     \undefined \def \showLCCN      #1{\unskip}     \fi
\ifx \shownote     \undefined \def \shownote      #1{#1}          \fi
\ifx \showarticletitle \undefined \def \showarticletitle #1{#1}   \fi
\ifx \showURL      \undefined \def \showURL       {\relax}        \fi
\providecommand\bibfield[2]{#2}
\providecommand\bibinfo[2]{#2}
\providecommand\natexlab[1]{#1}
\providecommand\showeprint[2][]{arXiv:#2}

\bibitem[Agapie et~al\mbox{.}(2022)]%
        {agapie_longitudinal_2022}
\bibfield{author}{\bibinfo{person}{Elena Agapie}, \bibinfo{person}{Patricia~A.
  Areán}, \bibinfo{person}{Gary Hsieh}, {and} \bibinfo{person}{Sean~A.
  Munson}.} \bibinfo{year}{2022}\natexlab{}.
\newblock \showarticletitle{A {Longitudinal} {Goal} {Setting} {Model} for
  {Addressing} {Complex} {Personal} {Problems} in {Mental} {Health}}.
\newblock \bibinfo{journal}{\emph{Proceedings of the ACM on Human-Computer
  Interaction}} \bibinfo{volume}{6}, \bibinfo{number}{CSCW2}
  (\bibinfo{date}{Nov.} \bibinfo{year}{2022}), \bibinfo{pages}{1--28}.
\newblock
\showISSN{2573-0142}
\urldef\tempurl%
\url{https://doi.org/10.1145/3555160}
\showDOI{\tempurl}


\bibitem[Baghaei et~al\mbox{.}(2020)]%
        {baghaei_time_2020}
\bibfield{author}{\bibinfo{person}{Nilufar Baghaei}, \bibinfo{person}{Lehan
  Stemmet}, \bibinfo{person}{Andrej Hlasnik}, \bibinfo{person}{Konstantin
  Emanov}, \bibinfo{person}{Sylvia Hach}, \bibinfo{person}{John~A. Naslund},
  \bibinfo{person}{Mark Billinghurst}, \bibinfo{person}{Imran Khaliq}, {and}
  \bibinfo{person}{Hai-Ning Liang}.} \bibinfo{year}{2020}\natexlab{}.
\newblock \showarticletitle{Time to {Get} {Personal}: {Individualised}
  {Virtual} {Reality} for {Mental} {Health}}. In
  \bibinfo{booktitle}{\emph{Extended {Abstracts} of the 2020 {CHI} {Conference}
  on {Human} {Factors} in {Computing} {Systems}}}. \bibinfo{publisher}{ACM},
  \bibinfo{address}{Honolulu HI USA}, \bibinfo{pages}{1--9}.
\newblock
\showISBNx{978-1-4503-6819-3}
\urldef\tempurl%
\url{https://doi.org/10.1145/3334480.3382932}
\showDOI{\tempurl}


\bibitem[Bakeman(2005)]%
        {bakeman_recommended_2005}
\bibfield{author}{\bibinfo{person}{Roger Bakeman}.}
  \bibinfo{year}{2005}\natexlab{}.
\newblock \showarticletitle{Recommended effect size statistics for repeated
  measures designs}.
\newblock \bibinfo{journal}{\emph{Behavior Research Methods}}
  \bibinfo{volume}{37}, \bibinfo{number}{3} (\bibinfo{date}{Aug.}
  \bibinfo{year}{2005}), \bibinfo{pages}{379--384}.
\newblock
\showISSN{1554-351X, 1554-3528}
\urldef\tempurl%
\url{https://doi.org/10.3758/BF03192707}
\showDOI{\tempurl}


\bibitem[Belloch et~al\mbox{.}(2004)]%
        {belloch_intrusive_2004}
\bibfield{author}{\bibinfo{person}{Amparo Belloch}, \bibinfo{person}{Carmen
  Morillo}, \bibinfo{person}{Mariela Lucero}, \bibinfo{person}{Elena Cabedo},
  {and} \bibinfo{person}{Carmen Carrió}.} \bibinfo{year}{2004}\natexlab{}.
\newblock \showarticletitle{Intrusive thoughts in non-clinical subjects: the
  role of frequency and unpleasantness on appraisal ratings and control
  strategies}.
\newblock \bibinfo{journal}{\emph{Clinical Psychology \& Psychotherapy}}
  \bibinfo{volume}{11}, \bibinfo{number}{2} (\bibinfo{date}{March}
  \bibinfo{year}{2004}), \bibinfo{pages}{100--110}.
\newblock
\showISSN{1063-3995, 1099-0879}
\urldef\tempurl%
\url{https://doi.org/10.1002/cpp.397}
\showDOI{\tempurl}


\bibitem[Berger et~al\mbox{.}(2022)]%
        {berger_appearance_2022}
\bibfield{author}{\bibinfo{person}{Jenny Berger}, \bibinfo{person}{Emmanuel
  Essah}, \bibinfo{person}{Tijana Blanusa}, {and} \bibinfo{person}{C.~Philip
  Beaman}.} \bibinfo{year}{2022}\natexlab{}.
\newblock \showarticletitle{The appearance of indoor plants and their effect on
  people's perceptions of indoor air quality and subjective well-being}.
\newblock \bibinfo{journal}{\emph{Building and Environment}}
  \bibinfo{volume}{219} (\bibinfo{date}{July} \bibinfo{year}{2022}),
  \bibinfo{pages}{109151}.
\newblock
\showISSN{03601323}
\urldef\tempurl%
\url{https://doi.org/10.1016/j.buildenv.2022.109151}
\showDOI{\tempurl}


\bibitem[Blandford et~al\mbox{.}(2016)]%
        {blandford_qualitative_2016}
\bibfield{author}{\bibinfo{person}{Ann Blandford}, \bibinfo{person}{Dominic
  Furniss}, {and} \bibinfo{person}{Stephann Makri}.}
  \bibinfo{year}{2016}\natexlab{}.
\newblock \bibinfo{booktitle}{\emph{Qualitative {HCI} {Research}: {Going}
  {Behind} the {Scenes}}}.
\newblock \bibinfo{publisher}{Springer International Publishing},
  \bibinfo{address}{Cham}.
\newblock
\showISBNx{978-3-031-01089-7 978-3-031-02217-3}
\urldef\tempurl%
\url{https://doi.org/10.1007/978-3-031-02217-3}
\showDOI{\tempurl}


\bibitem[Bosse et~al\mbox{.}(2013)]%
        {bosse_learning_2013}
\bibfield{author}{\bibinfo{person}{Tibor Bosse}, \bibinfo{person}{Charlotte
  Gerritsen}, \bibinfo{person}{Jeroen~De Man}, {and} \bibinfo{person}{Jan
  Treur}.} \bibinfo{year}{2013}\natexlab{}.
\newblock \showarticletitle{Learning {Emotion} {Regulation} {Strategies}: {A}
  {Cognitive} {Agent} {Model}}. In \bibinfo{booktitle}{\emph{2013
  {IEEE}/{WIC}/{ACM} {International} {Joint} {Conferences} on {Web}
  {Intelligence} ({WI}) and {Intelligent} {Agent} {Technologies} ({IAT})}}.
  \bibinfo{publisher}{IEEE}, \bibinfo{address}{Atlanta, GA, USA},
  \bibinfo{pages}{245--252}.
\newblock
\showISBNx{978-0-7695-5145-6 978-1-4799-2902-3}
\urldef\tempurl%
\url{https://doi.org/10.1109/WI-IAT.2013.116}
\showDOI{\tempurl}


\bibitem[Botella et~al\mbox{.}(2000)]%
        {botella_virtual_2000}
\bibfield{author}{\bibinfo{person}{Cristina Botella}, \bibinfo{person}{Rosa~M.
  Baños}, \bibinfo{person}{Helena Villa}, \bibinfo{person}{Conxa Perpiñá},
  {and} \bibinfo{person}{Azucena García-Palacios}.}
  \bibinfo{year}{2000}\natexlab{}.
\newblock \showarticletitle{Virtual reality in the treatment of claustrophobic
  fear: {A} controlled, multiple-baseline design}.
\newblock \bibinfo{journal}{\emph{Behavior Therapy}} \bibinfo{volume}{31},
  \bibinfo{number}{3} (\bibinfo{year}{2000}), \bibinfo{pages}{583--595}.
\newblock
\showISSN{00057894}
\urldef\tempurl%
\url{https://doi.org/10.1016/S0005-7894(00)80032-5}
\showDOI{\tempurl}


\bibitem[Briñol et~al\mbox{.}(2013)]%
        {brinol_treating_2013}
\bibfield{author}{\bibinfo{person}{Pablo Briñol}, \bibinfo{person}{Margarita
  Gascó}, \bibinfo{person}{Richard~E. Petty}, {and} \bibinfo{person}{Javier
  Horcajo}.} \bibinfo{year}{2013}\natexlab{}.
\newblock \showarticletitle{Treating {Thoughts} as {Material} {Objects} {Can}
  {Increase} or {Decrease} {Their} {Impact} on {Evaluation}}.
\newblock \bibinfo{journal}{\emph{Psychological Science}} \bibinfo{volume}{24},
  \bibinfo{number}{1} (\bibinfo{date}{Jan.} \bibinfo{year}{2013}),
  \bibinfo{pages}{41--47}.
\newblock
\showISSN{0956-7976, 1467-9280}
\urldef\tempurl%
\url{https://doi.org/10.1177/0956797612449176}
\showDOI{\tempurl}


\bibitem[Briñol et~al\mbox{.}(2017)]%
        {brinol_objectification_2017}
\bibfield{author}{\bibinfo{person}{Pablo Briñol}, \bibinfo{person}{Richard~E.
  Petty}, {and} \bibinfo{person}{Jennifer Belding}.}
  \bibinfo{year}{2017}\natexlab{}.
\newblock \showarticletitle{Objectification of people and thoughts: {An}
  attitude change perspective}.
\newblock \bibinfo{journal}{\emph{British Journal of Social Psychology}}
  \bibinfo{volume}{56}, \bibinfo{number}{2} (\bibinfo{date}{June}
  \bibinfo{year}{2017}), \bibinfo{pages}{233--249}.
\newblock
\showISSN{0144-6665, 2044-8309}
\urldef\tempurl%
\url{https://doi.org/10.1111/bjso.12183}
\showDOI{\tempurl}


\bibitem[Bryson(1995)]%
        {bryson_approaches_1995}
\bibfield{author}{\bibinfo{person}{Steve Bryson}.}
  \bibinfo{year}{1995}\natexlab{}.
\newblock \showarticletitle{Approaches to the successful design and
  implementation of {VR} applications}.
\newblock \bibinfo{journal}{\emph{Virtual reality applications}}
  (\bibinfo{year}{1995}), \bibinfo{pages}{3--15}.
\newblock


\bibitem[Butler et~al\mbox{.}(2006)]%
        {butler_empirical_2006}
\bibfield{author}{\bibinfo{person}{A Butler}, \bibinfo{person}{J Chapman},
  \bibinfo{person}{E Forman}, {and} \bibinfo{person}{A Beck}.}
  \bibinfo{year}{2006}\natexlab{}.
\newblock \showarticletitle{The empirical status of cognitive-behavioral
  therapy: {A} review of meta-analyses}.
\newblock \bibinfo{journal}{\emph{Clinical Psychology Review}}
  \bibinfo{volume}{26}, \bibinfo{number}{1} (\bibinfo{date}{Jan.}
  \bibinfo{year}{2006}), \bibinfo{pages}{17--31}.
\newblock
\showISSN{02727358}
\urldef\tempurl%
\url{https://doi.org/10.1016/j.cpr.2005.07.003}
\showDOI{\tempurl}


\bibitem[Carroll(2003)]%
        {carroll_hci_2003}
\bibfield{author}{\bibinfo{person}{John~M. Carroll}.}
  \bibinfo{year}{2003}\natexlab{}.
\newblock \bibinfo{booktitle}{\emph{{HCI} {Models}, {Theories}, and
  {Frameworks}: {Toward} a {Multidisciplinary} {Science}}}.
\newblock \bibinfo{publisher}{Elsevier}.
\newblock
\showISBNx{978-0-08-049141-7}


\bibitem[Clark and Rhyno(2005)]%
        {clark_unwanted_2005}
\bibfield{author}{\bibinfo{person}{David~A Clark} {and}
  \bibinfo{person}{Shelley Rhyno}.} \bibinfo{year}{2005}\natexlab{}.
\newblock \showarticletitle{Unwanted {Intrusive} {Thoughts} in {Nonclinical}
  {Individuals}: {Implications} for {Clinical} {Disorders}.}
\newblock  (\bibinfo{year}{2005}).
\newblock
\newblock
\shownote{Publisher: The Guilford Press}.


\bibitem[Cohen(2013)]%
        {cohen_statistical_2013}
\bibfield{author}{\bibinfo{person}{Jacob Cohen}.}
  \bibinfo{year}{2013}\natexlab{}.
\newblock \bibinfo{booktitle}{\emph{Statistical {Power} {Analysis} for the
  {Behavioral} {Sciences}} (\bibinfo{edition}{0} ed.)}.
\newblock \bibinfo{publisher}{Routledge}.
\newblock
\showISBNx{978-1-134-74270-7}
\urldef\tempurl%
\url{https://doi.org/10.4324/9780203771587}
\showDOI{\tempurl}


\bibitem[Dinas et~al\mbox{.}(2011)]%
        {dinas_effects_2011}
\bibfield{author}{\bibinfo{person}{P.~C. Dinas}, \bibinfo{person}{Y.
  Koutedakis}, {and} \bibinfo{person}{A.~D. Flouris}.}
  \bibinfo{year}{2011}\natexlab{}.
\newblock \showarticletitle{Effects of exercise and physical activity on
  depression}.
\newblock \bibinfo{journal}{\emph{Irish Journal of Medical Science}}
  \bibinfo{volume}{180}, \bibinfo{number}{2} (\bibinfo{date}{June}
  \bibinfo{year}{2011}), \bibinfo{pages}{319--325}.
\newblock
\showISSN{0021-1265, 1863-4362}
\urldef\tempurl%
\url{https://doi.org/10.1007/s11845-010-0633-9}
\showDOI{\tempurl}


\bibitem[Eikey et~al\mbox{.}(2021)]%
        {eikey_beyond_2021}
\bibfield{author}{\bibinfo{person}{Elizabeth~Victoria Eikey},
  \bibinfo{person}{Clara~Marques Caldeira}, \bibinfo{person}{Mayara~Costa
  Figueiredo}, \bibinfo{person}{Yunan Chen}, \bibinfo{person}{Jessica~L.
  Borelli}, \bibinfo{person}{Melissa Mazmanian}, {and} \bibinfo{person}{Kai
  Zheng}.} \bibinfo{year}{2021}\natexlab{}.
\newblock \showarticletitle{Beyond self-reflection: introducing the concept of
  rumination in personal informatics}.
\newblock \bibinfo{journal}{\emph{Personal and Ubiquitous Computing}}
  \bibinfo{volume}{25}, \bibinfo{number}{3} (\bibinfo{date}{June}
  \bibinfo{year}{2021}), \bibinfo{pages}{601--616}.
\newblock
\showISSN{1617-4909, 1617-4917}
\urldef\tempurl%
\url{https://doi.org/10.1007/s00779-021-01573-w}
\showDOI{\tempurl}


\bibitem[Elkin et~al\mbox{.}(2021)]%
        {elkin_aligned_2021}
\bibfield{author}{\bibinfo{person}{Lisa~A. Elkin}, \bibinfo{person}{Matthew
  Kay}, \bibinfo{person}{James~J. Higgins}, {and} \bibinfo{person}{Jacob~O.
  Wobbrock}.} \bibinfo{year}{2021}\natexlab{}.
\newblock \showarticletitle{An {Aligned} {Rank} {Transform} {Procedure} for
  {Multifactor} {Contrast} {Tests}}. In \bibinfo{booktitle}{\emph{The 34th
  {Annual} {ACM} {Symposium} on {User} {Interface} {Software} and
  {Technology}}}. \bibinfo{publisher}{ACM}, \bibinfo{address}{Virtual Event
  USA}, \bibinfo{pages}{754--768}.
\newblock
\showISBNx{978-1-4503-8635-7}
\urldef\tempurl%
\url{https://doi.org/10.1145/3472749.3474784}
\showDOI{\tempurl}


\bibitem[Evans(2006)]%
        {evans_child_2006}
\bibfield{author}{\bibinfo{person}{Gary~W. Evans}.}
  \bibinfo{year}{2006}\natexlab{}.
\newblock \showarticletitle{Child {Development} and the {Physical}
  {Environment}}.
\newblock \bibinfo{journal}{\emph{Annual Review of Psychology}}
  \bibinfo{volume}{57}, \bibinfo{number}{1} (\bibinfo{date}{Jan.}
  \bibinfo{year}{2006}), \bibinfo{pages}{423--451}.
\newblock
\showISSN{0066-4308, 1545-2085}
\urldef\tempurl%
\url{https://doi.org/10.1146/annurev.psych.57.102904.190057}
\showDOI{\tempurl}


\bibitem[Garcia-Palacios et~al\mbox{.}(2002)]%
        {garcia-palacios_virtual_2002}
\bibfield{author}{\bibinfo{person}{A Garcia-Palacios}, \bibinfo{person}{H
  Hoffman}, \bibinfo{person}{A Carlin}, \bibinfo{person}{T.A Furness}, {and}
  \bibinfo{person}{C Botella}.} \bibinfo{year}{2002}\natexlab{}.
\newblock \showarticletitle{Virtual reality in the treatment of spider phobia:
  a controlled study}.
\newblock \bibinfo{journal}{\emph{Behaviour Research and Therapy}}
  \bibinfo{volume}{40}, \bibinfo{number}{9} (\bibinfo{date}{Sept.}
  \bibinfo{year}{2002}), \bibinfo{pages}{983--993}.
\newblock
\showISSN{00057967}
\urldef\tempurl%
\url{https://doi.org/10.1016/S0005-7967(01)00068-7}
\showDOI{\tempurl}


\bibitem[Garnefski et~al\mbox{.}(2002)]%
        {garnefski_relationship_2002}
\bibfield{author}{\bibinfo{person}{Nadia Garnefski}, \bibinfo{person}{Tessa Van
  Den~Kommer}, \bibinfo{person}{Vivian Kraaij}, \bibinfo{person}{Jan Teerds},
  \bibinfo{person}{Jeroen Legerstee}, {and} \bibinfo{person}{Evert Onstein}.}
  \bibinfo{year}{2002}\natexlab{}.
\newblock \showarticletitle{The relationship between cognitive emotion
  regulation strategies and emotional problems: comparison between a clinical
  and a non‐clinical sample}.
\newblock \bibinfo{journal}{\emph{European Journal of Personality}}
  \bibinfo{volume}{16}, \bibinfo{number}{5} (\bibinfo{date}{Sept.}
  \bibinfo{year}{2002}), \bibinfo{pages}{403--420}.
\newblock
\showISSN{0890-2070, 1099-0984}
\urldef\tempurl%
\url{https://doi.org/10.1002/per.458}
\showDOI{\tempurl}


\bibitem[Gentile et~al\mbox{.}(2023)]%
        {gentile_nature_2023}
\bibfield{author}{\bibinfo{person}{Ambra Gentile}, \bibinfo{person}{Salvatore
  Ficarra}, \bibinfo{person}{Ewan Thomas}, \bibinfo{person}{Antonino Bianco},
  {and} \bibinfo{person}{Anna Nordstrom}.} \bibinfo{year}{2023}\natexlab{}.
\newblock \showarticletitle{Nature through virtual reality as a
  stress-reduction tool: {A} systematic review.}
\newblock \bibinfo{journal}{\emph{International Journal of Stress Management}}
  (\bibinfo{date}{July} \bibinfo{year}{2023}).
\newblock
\showISSN{1573-3424, 1072-5245}
\urldef\tempurl%
\url{https://doi.org/10.1037/str0000300}
\showDOI{\tempurl}


\bibitem[Greene et~al\mbox{.}(1989)]%
        {greene_toward_1989}
\bibfield{author}{\bibinfo{person}{Jennifer~C. Greene},
  \bibinfo{person}{Valerie~J. Caracelli}, {and} \bibinfo{person}{Wendy~F.
  Graham}.} \bibinfo{year}{1989}\natexlab{}.
\newblock \showarticletitle{Toward a {Conceptual} {Framework} for
  {Mixed}-{Method} {Evaluation} {Designs}}.
\newblock \bibinfo{journal}{\emph{Educational Evaluation and Policy Analysis}}
  \bibinfo{volume}{11}, \bibinfo{number}{3} (\bibinfo{date}{Sept.}
  \bibinfo{year}{1989}), \bibinfo{pages}{255--274}.
\newblock
\showISSN{0162-3737, 1935-1062}
\urldef\tempurl%
\url{https://doi.org/10.3102/01623737011003255}
\showDOI{\tempurl}


\bibitem[Grieger et~al\mbox{.}(2021)]%
        {grieger_trash_2021}
\bibfield{author}{\bibinfo{person}{Florian Grieger}, \bibinfo{person}{Holger
  Klapperich}, {and} \bibinfo{person}{Marc Hassenzahl}.}
  \bibinfo{year}{2021}\natexlab{}.
\newblock \showarticletitle{Trash {It}, {Punch} {It}, {Burn} {It} – {Using}
  {Virtual} {Reality} to {Support} {Coping} with {Negative} {Thoughts}}. In
  \bibinfo{booktitle}{\emph{Extended {Abstracts} of the 2021 {CHI} {Conference}
  on {Human} {Factors} in {Computing} {Systems}}}. \bibinfo{publisher}{ACM},
  \bibinfo{address}{Yokohama Japan}, \bibinfo{pages}{1--6}.
\newblock
\showISBNx{978-1-4503-8095-9}
\urldef\tempurl%
\url{https://doi.org/10.1145/3411763.3451738}
\showDOI{\tempurl}


\bibitem[Gross(1998)]%
        {gross_emerging_1998}
\bibfield{author}{\bibinfo{person}{James~J. Gross}.}
  \bibinfo{year}{1998}\natexlab{}.
\newblock \showarticletitle{The {Emerging} {Field} of {Emotion} {Regulation}:
  {An} {Integrative} {Review}}.
\newblock \bibinfo{journal}{\emph{Review of General Psychology}}
  \bibinfo{volume}{2}, \bibinfo{number}{3} (\bibinfo{date}{Sept.}
  \bibinfo{year}{1998}), \bibinfo{pages}{271--299}.
\newblock
\showISSN{1089-2680, 1939-1552}
\urldef\tempurl%
\url{https://doi.org/10.1037/1089-2680.2.3.271}
\showDOI{\tempurl}


\bibitem[Gross(2015)]%
        {gross_emotion_2015}
\bibfield{author}{\bibinfo{person}{James~J. Gross}.}
  \bibinfo{year}{2015}\natexlab{}.
\newblock \showarticletitle{Emotion {Regulation}: {Current} {Status} and
  {Future} {Prospects}}.
\newblock \bibinfo{journal}{\emph{Psychological Inquiry}} \bibinfo{volume}{26},
  \bibinfo{number}{1} (\bibinfo{date}{Jan.} \bibinfo{year}{2015}),
  \bibinfo{pages}{1--26}.
\newblock
\showISSN{1047-840X, 1532-7965}
\urldef\tempurl%
\url{https://doi.org/10.1080/1047840X.2014.940781}
\showDOI{\tempurl}


\bibitem[Gross and John(2003)]%
        {gross_individual_2003}
\bibfield{author}{\bibinfo{person}{James~J. Gross} {and}
  \bibinfo{person}{Oliver~P. John}.} \bibinfo{year}{2003}\natexlab{}.
\newblock \showarticletitle{Individual differences in two emotion regulation
  processes: {Implications} for affect, relationships, and well-being.}
\newblock \bibinfo{journal}{\emph{Journal of Personality and Social
  Psychology}} \bibinfo{volume}{85}, \bibinfo{number}{2}
  (\bibinfo{year}{2003}), \bibinfo{pages}{348--362}.
\newblock
\showISSN{1939-1315, 0022-3514}
\urldef\tempurl%
\url{https://doi.org/10.1037/0022-3514.85.2.348}
\showDOI{\tempurl}


\bibitem[Hacmun et~al\mbox{.}(2018)]%
        {hacmun_principles_2018}
\bibfield{author}{\bibinfo{person}{Irit Hacmun}, \bibinfo{person}{Dafna Regev},
  {and} \bibinfo{person}{Roy Salomon}.} \bibinfo{year}{2018}\natexlab{}.
\newblock \showarticletitle{The {Principles} of {Art} {Therapy} in {Virtual}
  {Reality}}.
\newblock \bibinfo{journal}{\emph{Frontiers in Psychology}}
  \bibinfo{volume}{9} (\bibinfo{date}{Oct.} \bibinfo{year}{2018}),
  \bibinfo{pages}{2082}.
\newblock
\showISSN{1664-1078}
\urldef\tempurl%
\url{https://doi.org/10.3389/fpsyg.2018.02082}
\showDOI{\tempurl}


\bibitem[Hacmun et~al\mbox{.}(2021)]%
        {hacmun_artistic_2021}
\bibfield{author}{\bibinfo{person}{Irit Hacmun}, \bibinfo{person}{Dafna Regev},
  {and} \bibinfo{person}{Roy Salomon}.} \bibinfo{year}{2021}\natexlab{}.
\newblock \showarticletitle{Artistic creation in virtual reality for art
  therapy: {A} qualitative study with expert art therapists}.
\newblock \bibinfo{journal}{\emph{The Arts in Psychotherapy}}
  \bibinfo{volume}{72} (\bibinfo{date}{Feb.} \bibinfo{year}{2021}),
  \bibinfo{pages}{101745}.
\newblock
\showISSN{01974556}
\urldef\tempurl%
\url{https://doi.org/10.1016/j.aip.2020.101745}
\showDOI{\tempurl}


\bibitem[Hanley et~al\mbox{.}(2015)]%
        {hanley_washing_2015}
\bibfield{author}{\bibinfo{person}{Adam~W. Hanley}, \bibinfo{person}{Alia~R.
  Warner}, \bibinfo{person}{Vincent~M. Dehili}, \bibinfo{person}{Angela~I.
  Canto}, {and} \bibinfo{person}{Eric~L. Garland}.}
  \bibinfo{year}{2015}\natexlab{}.
\newblock \showarticletitle{Washing {Dishes} to {Wash} the {Dishes}: {Brief}
  {Instruction} in an {Informal} {Mindfulness} {Practice}}.
\newblock \bibinfo{journal}{\emph{Mindfulness}} \bibinfo{volume}{6},
  \bibinfo{number}{5} (\bibinfo{date}{Oct.} \bibinfo{year}{2015}),
  \bibinfo{pages}{1095--1103}.
\newblock
\showISSN{1868-8527, 1868-8535}
\urldef\tempurl%
\url{https://doi.org/10.1007/s12671-014-0360-9}
\showDOI{\tempurl}


\bibitem[Hofmann and Asmundson(2008)]%
        {hofmann_acceptance_2008}
\bibfield{author}{\bibinfo{person}{Stefan~G. Hofmann} {and}
  \bibinfo{person}{Gordon~J.G. Asmundson}.} \bibinfo{year}{2008}\natexlab{}.
\newblock \showarticletitle{Acceptance and mindfulness-based therapy: {New}
  wave or old hat?}
\newblock \bibinfo{journal}{\emph{Clinical Psychology Review}}
  \bibinfo{volume}{28}, \bibinfo{number}{1} (\bibinfo{date}{Jan.}
  \bibinfo{year}{2008}), \bibinfo{pages}{1--16}.
\newblock
\showISSN{02727358}
\urldef\tempurl%
\url{https://doi.org/10.1016/j.cpr.2007.09.003}
\showDOI{\tempurl}


\bibitem[Hurtienne(2017)]%
        {hurtienne_how_2017}
\bibfield{author}{\bibinfo{person}{Jörn Hurtienne}.}
  \bibinfo{year}{2017}\natexlab{}.
\newblock \showarticletitle{How {Cognitive} {Linguistics} {Inspires} {HCI}:
  {Image} {Schemas} and {Image}-{Schematic} {Metaphors}}.
\newblock \bibinfo{journal}{\emph{International Journal of Human–Computer
  Interaction}} \bibinfo{volume}{33}, \bibinfo{number}{1} (\bibinfo{date}{Jan.}
  \bibinfo{year}{2017}), \bibinfo{pages}{1--20}.
\newblock
\showISSN{1044-7318, 1532-7590}
\urldef\tempurl%
\url{https://doi.org/10.1080/10447318.2016.1232227}
\showDOI{\tempurl}


\bibitem[Hurtienne and Meschke(2016)]%
        {hurtienne_soft_2016}
\bibfield{author}{\bibinfo{person}{Jörn Hurtienne} {and}
  \bibinfo{person}{Oliver Meschke}.} \bibinfo{year}{2016}\natexlab{}.
\newblock \showarticletitle{Soft {Pillows} and the {Near} and {Dear}:
  {Physical}-to-{Abstract} {Mappings} with {Image}-{Schematic} {Metaphors}}. In
  \bibinfo{booktitle}{\emph{Proceedings of the {TEI} '16: {Tenth}
  {International} {Conference} on {Tangible}, {Embedded}, and {Embodied}
  {Interaction}}}. \bibinfo{publisher}{ACM}, \bibinfo{address}{Eindhoven
  Netherlands}, \bibinfo{pages}{324--331}.
\newblock
\showISBNx{978-1-4503-3582-9}
\urldef\tempurl%
\url{https://doi.org/10.1145/2839462.2839483}
\showDOI{\tempurl}


\bibitem[Hurtienne et~al\mbox{.}(2009)]%
        {hurtienne_sad_2009}
\bibfield{author}{\bibinfo{person}{Jörn Hurtienne}, \bibinfo{person}{Christian
  Stößel}, {and} \bibinfo{person}{Katharina Weber}.}
  \bibinfo{year}{2009}\natexlab{}.
\newblock \showarticletitle{Sad is heavy and happy is light: population
  stereotypes of tangible object attributes}. In
  \bibinfo{booktitle}{\emph{Proceedings of the 3rd {International} {Conference}
  on {Tangible} and {Embedded} {Interaction}}}. \bibinfo{publisher}{ACM},
  \bibinfo{address}{Cambridge United Kingdom}, \bibinfo{pages}{61--68}.
\newblock
\showISBNx{978-1-60558-493-5}
\urldef\tempurl%
\url{https://doi.org/10.1145/1517664.1517686}
\showDOI{\tempurl}


\bibitem[Jiang and Ahmadpour(2021)]%
        {jiang_beyond_2021}
\bibfield{author}{\bibinfo{person}{Jade Jiang} {and} \bibinfo{person}{Naseem
  Ahmadpour}.} \bibinfo{year}{2021}\natexlab{}.
\newblock \showarticletitle{Beyond {Immersion}: {Designing} for {Reflection} in
  {Virtual} {Reality}}. In \bibinfo{booktitle}{\emph{33rd {Australian}
  {Conference} on {Human}-{Computer} {Interaction}}}. \bibinfo{publisher}{ACM},
  \bibinfo{address}{Melbourne VIC Australia}, \bibinfo{pages}{208--220}.
\newblock
\showISBNx{978-1-4503-9598-4}
\urldef\tempurl%
\url{https://doi.org/10.1145/3520495.3520501}
\showDOI{\tempurl}


\bibitem[Jung et~al\mbox{.}(2017)]%
        {jung_metaphors_2017}
\bibfield{author}{\bibinfo{person}{Heekyoung Jung}, \bibinfo{person}{Heather
  Wiltse}, \bibinfo{person}{Mikael Wiberg}, {and} \bibinfo{person}{Erik
  Stolterman}.} \bibinfo{year}{2017}\natexlab{}.
\newblock \showarticletitle{Metaphors, materialities, and affordances: {Hybrid}
  morphologies in the design of interactive artifacts}.
\newblock \bibinfo{journal}{\emph{Design Studies}}  \bibinfo{volume}{53}
  (\bibinfo{date}{Nov.} \bibinfo{year}{2017}), \bibinfo{pages}{24--46}.
\newblock
\showISSN{0142694X}
\urldef\tempurl%
\url{https://doi.org/10.1016/j.destud.2017.06.004}
\showDOI{\tempurl}


\bibitem[Kim and Maher(2019)]%
        {kim_metaphors_2019}
\bibfield{author}{\bibinfo{person}{Jingoog Kim} {and} \bibinfo{person}{Mary~Lou
  Maher}.} \bibinfo{year}{2019}\natexlab{}.
\newblock \showarticletitle{Metaphors, {Signifiers}, {Affordances}, and
  {Modalities} for {Designing} {Mobile} and {Embodied} {Interactive}
  {Systems}}. In \bibinfo{booktitle}{\emph{Proceedings of the 31st {Australian}
  {Conference} on {Human}-{Computer}-{Interaction}}}. \bibinfo{publisher}{ACM},
  \bibinfo{address}{Fremantle WA Australia}, \bibinfo{pages}{542--545}.
\newblock
\showISBNx{978-1-4503-7696-9}
\urldef\tempurl%
\url{https://doi.org/10.1145/3369457.3369527}
\showDOI{\tempurl}


\bibitem[Kopp and Craw(1998)]%
        {kopp_metaphoric_1998}
\bibfield{author}{\bibinfo{person}{Richard~R. Kopp} {and}
  \bibinfo{person}{Michael~Jay Craw}.} \bibinfo{year}{1998}\natexlab{}.
\newblock \showarticletitle{Metaphoric language, metaphoric cognition, and
  cognitive therapy.}
\newblock \bibinfo{journal}{\emph{Psychotherapy: Theory, Research, Practice,
  Training}} \bibinfo{volume}{35}, \bibinfo{number}{3} (\bibinfo{year}{1998}),
  \bibinfo{pages}{306--311}.
\newblock
\showISSN{1939-1536, 0033-3204}
\urldef\tempurl%
\url{https://doi.org/10.1037/h0087795}
\showDOI{\tempurl}


\bibitem[Lang et~al\mbox{.}(2015)]%
        {lang_effects_2015}
\bibfield{author}{\bibinfo{person}{Martin Lang}, \bibinfo{person}{Jan
  Krátký}, \bibinfo{person}{John~H. Shaver}, \bibinfo{person}{Danijela
  Jerotijević}, {and} \bibinfo{person}{Dimitris Xygalatas}.}
  \bibinfo{year}{2015}\natexlab{}.
\newblock \showarticletitle{Effects of {Anxiety} on {Spontaneous} {Ritualized}
  {Behavior}}.
\newblock \bibinfo{journal}{\emph{Current Biology}} \bibinfo{volume}{25},
  \bibinfo{number}{14} (\bibinfo{date}{July} \bibinfo{year}{2015}),
  \bibinfo{pages}{1892--1897}.
\newblock
\showISSN{09609822}
\urldef\tempurl%
\url{https://doi.org/10.1016/j.cub.2015.05.049}
\showDOI{\tempurl}


\bibitem[Langer(2009)]%
        {langer_philosophy_2009}
\bibfield{author}{\bibinfo{person}{Susanne~K. Langer}.}
  \bibinfo{year}{2009}\natexlab{}.
\newblock \bibinfo{booktitle}{\emph{Philosophy in a {New} {Key}: {A} {Study} in
  the {Symbolism} of {Reason}, {Rite}, and {Art}, {Third} {Edition}}}.
\newblock \bibinfo{publisher}{Harvard University Press}.
\newblock
\showISBNx{978-0-674-03994-0}


\bibitem[Larsson et~al\mbox{.}(2016)]%
        {larsson_using_2016}
\bibfield{author}{\bibinfo{person}{Andreas Larsson}, \bibinfo{person}{Nic
  Hooper}, \bibinfo{person}{Lisa~A. Osborne}, \bibinfo{person}{Paul Bennett},
  {and} \bibinfo{person}{Louise McHugh}.} \bibinfo{year}{2016}\natexlab{}.
\newblock \showarticletitle{Using {Brief} {Cognitive} {Restructuring} and
  {Cognitive} {Defusion} {Techniques} to {Cope} {With} {Negative} {Thoughts}}.
\newblock \bibinfo{journal}{\emph{Behavior Modification}} \bibinfo{volume}{40},
  \bibinfo{number}{3} (\bibinfo{date}{May} \bibinfo{year}{2016}),
  \bibinfo{pages}{452--482}.
\newblock
\showISSN{0145-4455, 1552-4167}
\urldef\tempurl%
\url{https://doi.org/10.1177/0145445515621488}
\showDOI{\tempurl}


\bibitem[Lee and Schwarz(2011)]%
        {lee_wiping_2011}
\bibfield{author}{\bibinfo{person}{Spike W.~S. Lee} {and}
  \bibinfo{person}{Norbert Schwarz}.} \bibinfo{year}{2011}\natexlab{}.
\newblock \showarticletitle{Wiping the {Slate} {Clean}: {Psychological}
  {Consequences} of {Physical} {Cleansing}}.
\newblock \bibinfo{journal}{\emph{Current Directions in Psychological Science}}
  \bibinfo{volume}{20}, \bibinfo{number}{5} (\bibinfo{date}{Oct.}
  \bibinfo{year}{2011}), \bibinfo{pages}{307--311}.
\newblock
\showISSN{0963-7214, 1467-8721}
\urldef\tempurl%
\url{https://doi.org/10.1177/0963721411422694}
\showDOI{\tempurl}


\bibitem[Loomis et~al\mbox{.}(1999)]%
        {loomis_immersive_1999}
\bibfield{author}{\bibinfo{person}{Jack~M. Loomis}, \bibinfo{person}{James~J.
  Blascovich}, {and} \bibinfo{person}{Andrew~C. Beall}.}
  \bibinfo{year}{1999}\natexlab{}.
\newblock \showarticletitle{Immersive virtual environment technology as a basic
  research tool in psychology}.
\newblock \bibinfo{journal}{\emph{Behavior Research Methods, Instruments, \&
  Computers}} \bibinfo{volume}{31}, \bibinfo{number}{4} (\bibinfo{date}{Dec.}
  \bibinfo{year}{1999}), \bibinfo{pages}{557--564}.
\newblock
\showISSN{0743-3808, 1532-5970}
\urldef\tempurl%
\url{https://doi.org/10.3758/BF03200735}
\showDOI{\tempurl}


\bibitem[Lovell and Richards(2000)]%
        {lovell_multiple_2000}
\bibfield{author}{\bibinfo{person}{Karina Lovell} {and} \bibinfo{person}{David
  Richards}.} \bibinfo{year}{2000}\natexlab{}.
\newblock \showarticletitle{Multiple access points and levels of entry
  ({MAPLE}): {Ensuring} choice, accessibility and equity for {CBT} services.}
\newblock \bibinfo{journal}{\emph{Behavioural and Cognitive Psychotherapy}}
  \bibinfo{volume}{28}, \bibinfo{number}{4} (\bibinfo{date}{Oct.}
  \bibinfo{year}{2000}), \bibinfo{pages}{379--391}.
\newblock
\showISSN{1352-4658, 1469-1833}
\urldef\tempurl%
\url{https://doi.org/10.1017/S1352465800004070}
\showDOI{\tempurl}


\bibitem[McRae and Gross(2020)]%
        {mcrae_emotion_2020}
\bibfield{author}{\bibinfo{person}{Kateri McRae} {and}
  \bibinfo{person}{James~J. Gross}.} \bibinfo{year}{2020}\natexlab{}.
\newblock \showarticletitle{Emotion regulation.}
\newblock \bibinfo{journal}{\emph{Emotion}} \bibinfo{volume}{20},
  \bibinfo{number}{1} (\bibinfo{date}{Feb.} \bibinfo{year}{2020}),
  \bibinfo{pages}{1--9}.
\newblock
\showISSN{1931-1516, 1528-3542}
\urldef\tempurl%
\url{https://doi.org/10.1037/emo0000703}
\showDOI{\tempurl}


\bibitem[Mostajeran et~al\mbox{.}(2023)]%
        {mostajeran_adding_2023}
\bibfield{author}{\bibinfo{person}{Fariba Mostajeran}, \bibinfo{person}{Frank
  Steinicke}, \bibinfo{person}{Sarah Reinhart}, \bibinfo{person}{Wolfgang
  Stuerzlinger}, \bibinfo{person}{Bernhard~E. Riecke}, {and}
  \bibinfo{person}{Simone Kühn}.} \bibinfo{year}{2023}\natexlab{}.
\newblock \showarticletitle{Adding virtual plants leads to higher cognitive
  performance and psychological well-being in virtual reality}.
\newblock \bibinfo{journal}{\emph{Scientific Reports}} \bibinfo{volume}{13},
  \bibinfo{number}{1} (\bibinfo{date}{May} \bibinfo{year}{2023}),
  \bibinfo{pages}{8053}.
\newblock
\showISSN{2045-2322}
\urldef\tempurl%
\url{https://doi.org/10.1038/s41598-023-34718-3}
\showDOI{\tempurl}


\bibitem[Navinés et~al\mbox{.}(2008)]%
        {navines_interaction_2008}
\bibfield{author}{\bibinfo{person}{Ricard Navinés}, \bibinfo{person}{Rocío
  Martín-Santos}, \bibinfo{person}{Esther Gómez-Gil},
  \bibinfo{person}{María~J. Martínez De~Osaba}, {and}
  \bibinfo{person}{Cristòbal Gastó}.} \bibinfo{year}{2008}\natexlab{}.
\newblock \showarticletitle{Interaction between serotonin 5-{HT1A} receptors
  and beta-endorphins modulates antidepressant response}.
\newblock \bibinfo{journal}{\emph{Progress in Neuro-Psychopharmacology and
  Biological Psychiatry}} \bibinfo{volume}{32}, \bibinfo{number}{8}
  (\bibinfo{date}{Dec.} \bibinfo{year}{2008}), \bibinfo{pages}{1804--1809}.
\newblock
\showISSN{02785846}
\urldef\tempurl%
\url{https://doi.org/10.1016/j.pnpbp.2008.07.021}
\showDOI{\tempurl}


\bibitem[Otto(2000)]%
        {otto_stories_2000}
\bibfield{author}{\bibinfo{person}{Michael~W. Otto}.}
  \bibinfo{year}{2000}\natexlab{}.
\newblock \showarticletitle{Stories and metaphors in cognitive-behavior
  therapy}.
\newblock \bibinfo{journal}{\emph{Cognitive and Behavioral Practice}}
  \bibinfo{volume}{7}, \bibinfo{number}{2} (\bibinfo{date}{March}
  \bibinfo{year}{2000}), \bibinfo{pages}{166--172}.
\newblock
\showISSN{10777229}
\urldef\tempurl%
\url{https://doi.org/10.1016/S1077-7229(00)80027-9}
\showDOI{\tempurl}


\bibitem[Rothbaum et~al\mbox{.}(2000)]%
        {rothbaum_controlled_2000}
\bibfield{author}{\bibinfo{person}{Barbara~Olasov Rothbaum},
  \bibinfo{person}{Larry Hodges}, \bibinfo{person}{Samantha Smith},
  \bibinfo{person}{Jeong~Hwan Lee}, {and} \bibinfo{person}{Larry Price}.}
  \bibinfo{year}{2000}\natexlab{}.
\newblock \showarticletitle{A controlled study of virtual reality exposure
  therapy for the fear of flying.}
\newblock \bibinfo{journal}{\emph{Journal of Consulting and Clinical
  Psychology}} \bibinfo{volume}{68}, \bibinfo{number}{6} (\bibinfo{date}{Dec.}
  \bibinfo{year}{2000}), \bibinfo{pages}{1020--1026}.
\newblock
\showISSN{1939-2117, 0022-006X}
\urldef\tempurl%
\url{https://doi.org/10.1037/0022-006X.68.6.1020}
\showDOI{\tempurl}


\bibitem[Rozgonjuk and Elhai(2021)]%
        {rozgonjuk_emotion_2021}
\bibfield{author}{\bibinfo{person}{Dmitri Rozgonjuk} {and}
  \bibinfo{person}{Jon~D. Elhai}.} \bibinfo{year}{2021}\natexlab{}.
\newblock \showarticletitle{Emotion regulation in relation to smartphone use:
  {Process} smartphone use mediates the association between expressive
  suppression and problematic smartphone use}.
\newblock \bibinfo{journal}{\emph{Current Psychology}} \bibinfo{volume}{40},
  \bibinfo{number}{7} (\bibinfo{date}{July} \bibinfo{year}{2021}),
  \bibinfo{pages}{3246--3255}.
\newblock
\showISSN{1046-1310, 1936-4733}
\urldef\tempurl%
\url{https://doi.org/10.1007/s12144-019-00271-4}
\showDOI{\tempurl}


\bibitem[Saxbe and Repetti(2010)]%
        {saxbe_no_2010}
\bibfield{author}{\bibinfo{person}{Darby~E. Saxbe} {and} \bibinfo{person}{Rena
  Repetti}.} \bibinfo{year}{2010}\natexlab{}.
\newblock \showarticletitle{No {Place} {Like} {Home}: {Home} {Tours}
  {Correlate} {With} {Daily} {Patterns} of {Mood} and {Cortisol}}.
\newblock \bibinfo{journal}{\emph{Personality and Social Psychology Bulletin}}
  \bibinfo{volume}{36}, \bibinfo{number}{1} (\bibinfo{date}{Jan.}
  \bibinfo{year}{2010}), \bibinfo{pages}{71--81}.
\newblock
\showISSN{0146-1672, 1552-7433}
\urldef\tempurl%
\url{https://doi.org/10.1177/0146167209352864}
\showDOI{\tempurl}


\bibitem[Semsioglu et~al\mbox{.}(2021)]%
        {semsioglu_isles_2021}
\bibfield{author}{\bibinfo{person}{Sinem Semsioglu}, \bibinfo{person}{Pelin
  Karaturhan}, \bibinfo{person}{Saliha Akbas}, {and}
  \bibinfo{person}{Asim~Evren Yantac}.} \bibinfo{year}{2021}\natexlab{}.
\newblock \showarticletitle{Isles of {Emotion}: {Emotionally} {Expressive}
  {Social} {Virtual} {Spaces} for {Reflection} and {Communication}}. In
  \bibinfo{booktitle}{\emph{Creativity and {Cognition}}}.
  \bibinfo{publisher}{ACM}, \bibinfo{address}{Virtual Event Italy},
  \bibinfo{pages}{1--10}.
\newblock
\showISBNx{978-1-4503-8376-9}
\urldef\tempurl%
\url{https://doi.org/10.1145/3450741.3466805}
\showDOI{\tempurl}


\bibitem[Shi et~al\mbox{.}(2023)]%
        {shi_instant_2023}
\bibfield{author}{\bibinfo{person}{Yaoxi Shi}, \bibinfo{person}{Peter Koval},
  \bibinfo{person}{Vassilis Kostakos}, \bibinfo{person}{Jorge Goncalves}, {and}
  \bibinfo{person}{Greg Wadley}.} \bibinfo{year}{2023}\natexlab{}.
\newblock \showarticletitle{“{Instant} {Happiness}”: {Smartphones} as tools
  for everyday emotion regulation}.
\newblock \bibinfo{journal}{\emph{International Journal of Human-Computer
  Studies}}  \bibinfo{volume}{170} (\bibinfo{date}{Feb.} \bibinfo{year}{2023}),
  \bibinfo{pages}{102958}.
\newblock
\showISSN{10715819}
\urldef\tempurl%
\url{https://doi.org/10.1016/j.ijhcs.2022.102958}
\showDOI{\tempurl}


\bibitem[Shosha(2020)]%
        {shosha_brief_2020}
\bibfield{author}{\bibinfo{person}{Dr.~Mohammed Shosha}.}
  \bibinfo{year}{2020}\natexlab{}.
\newblock \showarticletitle{A brief introduction to therapeutic boxing}.
\newblock \bibinfo{journal}{\emph{International Journal of Physiology,
  Nutrition and Physical Education}} \bibinfo{volume}{5}, \bibinfo{number}{2}
  (\bibinfo{date}{July} \bibinfo{year}{2020}), \bibinfo{pages}{29--31}.
\newblock
\showISSN{24560057, 24560057}
\urldef\tempurl%
\url{https://doi.org/10.22271/journalofsport.2020.v5.i2a.1977}
\showDOI{\tempurl}


\bibitem[Slovak et~al\mbox{.}(2023)]%
        {slovak_designing_2023}
\bibfield{author}{\bibinfo{person}{Petr Slovak}, \bibinfo{person}{Alissa
  Antle}, \bibinfo{person}{Nikki Theofanopoulou}, \bibinfo{person}{Claudia
  Daudén~Roquet}, \bibinfo{person}{James Gross}, {and}
  \bibinfo{person}{Katherine Isbister}.} \bibinfo{year}{2023}\natexlab{}.
\newblock \showarticletitle{Designing for {Emotion} {Regulation}
  {Interventions}: {An} {Agenda} for {HCI} {Theory} and {Research}}.
\newblock \bibinfo{journal}{\emph{ACM Transactions on Computer-Human
  Interaction}} \bibinfo{volume}{30}, \bibinfo{number}{1} (\bibinfo{date}{Feb.}
  \bibinfo{year}{2023}), \bibinfo{pages}{1--51}.
\newblock
\showISSN{1073-0516, 1557-7325}
\urldef\tempurl%
\url{https://doi.org/10.1145/3569898}
\showDOI{\tempurl}


\bibitem[Smith et~al\mbox{.}(2022)]%
        {smith_digital_2022}
\bibfield{author}{\bibinfo{person}{Wally Smith}, \bibinfo{person}{Greg Wadley},
  \bibinfo{person}{Sarah Webber}, \bibinfo{person}{Benjamin Tag},
  \bibinfo{person}{Vassilis Kostakos}, \bibinfo{person}{Peter Koval}, {and}
  \bibinfo{person}{James~J. Gross}.} \bibinfo{year}{2022}\natexlab{}.
\newblock \showarticletitle{Digital {Emotion} {Regulation} in {Everyday}
  {Life}}. In \bibinfo{booktitle}{\emph{{CHI} {Conference} on {Human} {Factors}
  in {Computing} {Systems}}}. \bibinfo{publisher}{ACM}, \bibinfo{address}{New
  Orleans LA USA}, \bibinfo{pages}{1--15}.
\newblock
\showISBNx{978-1-4503-9157-3}
\urldef\tempurl%
\url{https://doi.org/10.1145/3491102.3517573}
\showDOI{\tempurl}


\bibitem[Steven~D. and Philip~C.(2011)]%
        {steven_d_automatic_2011}
\bibfield{author}{\bibinfo{person}{Hollon Steven~D.} {and}
  \bibinfo{person}{Kendall Philip~C.}} \bibinfo{year}{2011}\natexlab{}.
\newblock \bibinfo{title}{Automatic {Thoughts} {Questionnaire}}.
\newblock
\newblock
\urldef\tempurl%
\url{https://doi.org/10.1037/t00735-000}
\showDOI{\tempurl}


\bibitem[Titchener(1914)]%
        {titchener_text-book_1914}
\bibfield{author}{\bibinfo{person}{Edward~Bradford Titchener}.}
  \bibinfo{year}{1914}\natexlab{}.
\newblock \bibinfo{booktitle}{\emph{A text-book of psychology.}}
\newblock \bibinfo{publisher}{MacMillan Co}, \bibinfo{address}{New York}.
\newblock
\urldef\tempurl%
\url{https://doi.org/10.1037/10907-000}
\showDOI{\tempurl}


\bibitem[Wadley et~al\mbox{.}(2020)]%
        {wadley_digital_2020}
\bibfield{author}{\bibinfo{person}{Greg Wadley}, \bibinfo{person}{Wally Smith},
  \bibinfo{person}{Peter Koval}, {and} \bibinfo{person}{James~J. Gross}.}
  \bibinfo{year}{2020}\natexlab{}.
\newblock \showarticletitle{Digital {Emotion} {Regulation}}.
\newblock \bibinfo{journal}{\emph{Current Directions in Psychological Science}}
  \bibinfo{volume}{29}, \bibinfo{number}{4} (\bibinfo{date}{Aug.}
  \bibinfo{year}{2020}), \bibinfo{pages}{412--418}.
\newblock
\showISSN{0963-7214, 1467-8721}
\urldef\tempurl%
\url{https://doi.org/10.1177/0963721420920592}
\showDOI{\tempurl}


\bibitem[Wagener et~al\mbox{.}(2021)]%
        {wagener_role_2021}
\bibfield{author}{\bibinfo{person}{Nadine Wagener}, \bibinfo{person}{Tu~Dinh
  Duong}, \bibinfo{person}{Johannes Schöning}, \bibinfo{person}{Yvonne
  Rogers}, {and} \bibinfo{person}{Jasmin Niess}.}
  \bibinfo{year}{2021}\natexlab{}.
\newblock \showarticletitle{The {Role} of {Mobile} and {Virtual} {Reality}
  {Applications} to {Support} {Well}-{Being}: {An} {Expert} {View} and
  {Systematic} {App} {Review}}.
\newblock In \bibinfo{booktitle}{\emph{Human-{Computer} {Interaction} –
  {INTERACT} 2021}}, \bibfield{editor}{\bibinfo{person}{Carmelo Ardito},
  \bibinfo{person}{Rosa Lanzilotti}, \bibinfo{person}{Alessio Malizia},
  \bibinfo{person}{Helen Petrie}, \bibinfo{person}{Antonio Piccinno},
  \bibinfo{person}{Giuseppe Desolda}, {and} \bibinfo{person}{Kori Inkpen}}
  (Eds.). Vol.~\bibinfo{volume}{12935}. \bibinfo{publisher}{Springer
  International Publishing}, \bibinfo{address}{Cham},
  \bibinfo{pages}{262--283}.
\newblock
\showISBNx{978-3-030-85609-0 978-3-030-85610-6}
\urldef\tempurl%
\url{https://doi.org/10.1007/978-3-030-85610-6_16}
\showDOI{\tempurl}


\bibitem[Wagener et~al\mbox{.}(2022)]%
        {wagener_mood_2022}
\bibfield{author}{\bibinfo{person}{Nadine Wagener}, \bibinfo{person}{Jasmin
  Niess}, \bibinfo{person}{Yvonne Rogers}, {and} \bibinfo{person}{Johannes
  Schöning}.} \bibinfo{year}{2022}\natexlab{}.
\newblock \showarticletitle{Mood {Worlds}: {A} {Virtual} {Environment} for
  {Autonomous} {Emotional} {Expression}}. In \bibinfo{booktitle}{\emph{{CHI}
  {Conference} on {Human} {Factors} in {Computing} {Systems}}}.
  \bibinfo{publisher}{ACM}, \bibinfo{address}{New Orleans LA USA},
  \bibinfo{pages}{1--16}.
\newblock
\showISBNx{978-1-4503-9157-3}
\urldef\tempurl%
\url{https://doi.org/10.1145/3491102.3501861}
\showDOI{\tempurl}


\bibitem[Wagener et~al\mbox{.}(2023a)]%
        {wagener_selvreflect_2023}
\bibfield{author}{\bibinfo{person}{Nadine Wagener}, \bibinfo{person}{Leon
  Reicherts}, \bibinfo{person}{Nima Zargham}, \bibinfo{person}{Natalia
  Bartłomiejczyk}, \bibinfo{person}{Ava~Elizabeth Scott},
  \bibinfo{person}{Katherine Wang}, \bibinfo{person}{Marit Bentvelzen},
  \bibinfo{person}{Evropi Stefanidi}, \bibinfo{person}{Thomas Mildner},
  \bibinfo{person}{Yvonne Rogers}, {and} \bibinfo{person}{Jasmin Niess}.}
  \bibinfo{year}{2023}\natexlab{a}.
\newblock \showarticletitle{{SelVReflect}: {A} {Guided} {VR} {Experience}
  {Fostering} {Reflection} on {Personal} {Challenges}}. In
  \bibinfo{booktitle}{\emph{Proceedings of the 2023 {CHI} {Conference} on
  {Human} {Factors} in {Computing} {Systems}}}. \bibinfo{publisher}{ACM},
  \bibinfo{address}{Hamburg Germany}, \bibinfo{pages}{1--17}.
\newblock
\showISBNx{978-1-4503-9421-5}
\urldef\tempurl%
\url{https://doi.org/10.1145/3544548.3580763}
\showDOI{\tempurl}


\bibitem[Wagener et~al\mbox{.}(2023b)]%
        {wagener_letting_2023}
\bibfield{author}{\bibinfo{person}{Nadine Wagener}, \bibinfo{person}{Johannes
  Schöning}, \bibinfo{person}{Yvonne Rogers}, {and} \bibinfo{person}{Jasmin
  Niess}.} \bibinfo{year}{2023}\natexlab{b}.
\newblock \bibinfo{booktitle}{\emph{Letting {It} {Go}: {Four} {Design}
  {Concepts} to {Support} {Emotion} {Regulation} in {Virtual} {Reality}}}.
\newblock


\bibitem[Wang et~al\mbox{.}(2018)]%
        {wang_break_2018}
\bibfield{author}{\bibinfo{person}{Xinyue Wang}, \bibinfo{person}{Kelong Lu},
  \bibinfo{person}{Mark~A. Runco}, {and} \bibinfo{person}{Ning Hao}.}
  \bibinfo{year}{2018}\natexlab{}.
\newblock \showarticletitle{Break the “wall” and become creative:
  {Enacting} embodied metaphors in virtual reality}.
\newblock \bibinfo{journal}{\emph{Consciousness and Cognition}}
  \bibinfo{volume}{62} (\bibinfo{date}{July} \bibinfo{year}{2018}),
  \bibinfo{pages}{102--109}.
\newblock
\showISSN{10538100}
\urldef\tempurl%
\url{https://doi.org/10.1016/j.concog.2018.03.004}
\showDOI{\tempurl}


\bibitem[Webb et~al\mbox{.}(2012)]%
        {webb_dealing_2012}
\bibfield{author}{\bibinfo{person}{Thomas~L. Webb}, \bibinfo{person}{Eleanor
  Miles}, {and} \bibinfo{person}{Paschal Sheeran}.}
  \bibinfo{year}{2012}\natexlab{}.
\newblock \showarticletitle{Dealing with feeling: {A} meta-analysis of the
  effectiveness of strategies derived from the process model of emotion
  regulation.}
\newblock \bibinfo{journal}{\emph{Psychological Bulletin}}
  \bibinfo{volume}{138}, \bibinfo{number}{4} (\bibinfo{date}{July}
  \bibinfo{year}{2012}), \bibinfo{pages}{775--808}.
\newblock
\showISSN{1939-1455, 0033-2909}
\urldef\tempurl%
\url{https://doi.org/10.1037/a0027600}
\showDOI{\tempurl}


\bibitem[Wobbrock et~al\mbox{.}(2011)]%
        {wobbrock_aligned_2011}
\bibfield{author}{\bibinfo{person}{Jacob~O. Wobbrock}, \bibinfo{person}{Leah
  Findlater}, \bibinfo{person}{Darren Gergle}, {and} \bibinfo{person}{James~J.
  Higgins}.} \bibinfo{year}{2011}\natexlab{}.
\newblock \showarticletitle{The aligned rank transform for nonparametric
  factorial analyses using only anova procedures}. In
  \bibinfo{booktitle}{\emph{Proceedings of the {SIGCHI} {Conference} on {Human}
  {Factors} in {Computing} {Systems}}}. \bibinfo{publisher}{ACM},
  \bibinfo{address}{Vancouver BC Canada}, \bibinfo{pages}{143--146}.
\newblock
\showISBNx{978-1-4503-0228-9}
\urldef\tempurl%
\url{https://doi.org/10.1145/1978942.1978963}
\showDOI{\tempurl}


\bibitem[Yildirim et~al\mbox{.}(2011)]%
        {yildirim_effects_2011}
\bibfield{author}{\bibinfo{person}{Kemal Yildirim}, \bibinfo{person}{M.~Lutfi
  Hidayetoglu}, {and} \bibinfo{person}{Aysen Capanoglu}.}
  \bibinfo{year}{2011}\natexlab{}.
\newblock \showarticletitle{Effects of {Interior} {Colors} on {Mood} and
  {Preference}: {Comparisons} of {Two} {Living} {Rooms}}.
\newblock \bibinfo{journal}{\emph{Perceptual and Motor Skills}}
  \bibinfo{volume}{112}, \bibinfo{number}{2} (\bibinfo{date}{April}
  \bibinfo{year}{2011}), \bibinfo{pages}{509--524}.
\newblock
\showISSN{0031-5125, 1558-688X}
\urldef\tempurl%
\url{https://doi.org/10.2466/24.27.PMS.112.2.509-524}
\showDOI{\tempurl}


\end{thebibliography}
